\documentclass{aa}
\usepackage[usenames,dvipsnames]{xcolor}
\usepackage{graphicx}
\graphicspath{{./figs/}}
\usepackage{txfonts}
\usepackage{gensymb}
\usepackage{tikz}
\usepackage{courier}
\usepackage[bf,normalsize]{subfigure}
\usepackage{afterpage}
\usepackage{siunitx}
\usepackage{newtxtext}
\usepackage[frenchmath,varg]{newtxmath}
\usepackage{natbib,twoopt}
\usepackage[breaklinks=true,colorlinks=true,linkcolor=NavyBlue!50!Emerald,citecolor=Gray!70!ProcessBlue,filecolor=NavyBlue!50!Emerald,urlcolor=NavyBlue!50!Emerald]{hyperref}
\usepackage[para,online,flushleft]{threeparttable}
\usepackage{orcidlink}

\bibpunct{(}{)}{;}{a}{}{,} %% natbib format for A&A and ApJ
\begin{document} 

   \titlerunning{JOLINES}
   \authorrunning{Fern\'andez-Ontiveros et al.}
   \title{J-PAS: A value-added catalogue of optical line intensities for nebular emission galaxies (JOLINES)}
   % A value-added catalogue of emission-line fluxes for star-forming galaxies and AGN in J-PAS surveys

   %\subtitle{JOLINES}
   % \thanks{Based on VLT XX.X-XXXA, YY.Y-YYYYA}

   \author{J.A.\,Fernández-Ontiveros\orcidlink{0000-0001-9490-899X}\inst{\ref{CEFCA},\ref{CEFCA-UA}}\thanks{\email{\sf \href{mailto:j.a.fernandez.ontiveros@gmail.com}{j.a.fernandez.ontiveros@gmail.com}, \href{mailto:jafernandez@cefca.es}{jafernandez@cefca.es}}} \and C.\,López-Sanjuan\orcidlink{0000-0002-5743-3160}\inst{\ref{CEFCA},\ref{CEFCA-UA}} \and A.\,Hernán-Caballero\orcidlink{0000-0002-4237-5500}\inst{\ref{CEFCA},\ref{CEFCA-UA}} \and A.\,Lumbreras-Calle\orcidlink{0000-0002-6696-7834}\inst{\ref{CEFCA}} \and J.\,Iglesias-Páramo\orcidlink{0000-0003-2726-6370}\inst{\ref{IAA}} \and A.\,Torralba\orcidlink{0000-0001-5586-6950}\inst{\ref{ISTA}} \and R.M.\,González Delgado\orcidlink{https://orcid.org/0000-0003-1058-1577}\inst{\ref{IAA}} \and A.\,del\,Pino\inst{\ref{IAA}} \and Rahna\,P.T.\inst{\ref{CEFCA}} \and I.E.\,López\inst{\ref{INAF-OAB}} \and R.\,Amorín\inst{\ref{IAA}} \and  J.M.\,Vílchez\inst{\ref{IAA}} \and C.\,Kehrig\inst{\ref{IAA}} \and I.\,Breda\inst{{\ref{UNIVIE}}} \and D.\,Fernández\,Gil\inst{\ref{CEFCA}} \and F.D.\,Arizo-Borillo\inst{\ref{CEFCA}} \and A.\,Giménez-Alcázar\inst{\ref{IAA}} \and E.\,Pérez-Montero\inst{\ref{IAA}} \and F.J.\,Sáez\,Ruiz\inst{\ref{CEFCA}} \and N.\,Acharya\inst{\ref{CEFCA}} %\fnmsep
   % Founders
   R.\,Abramo\inst{\ref{USP}} \and J.\,Alcaniz\inst{\ref{ON}} \and N.\,Benítez\inst{\ref{IR}} \and S.\,Bonoli\inst{\ref{DIPC},\ref{CEFCA}} \and S.\,Carneiro\inst{\ref{ON}} \and J.\,Cenarro\inst{\ref{CEFCA},\ref{CEFCA-UA}} \and D.\,Cristóbal-Hornillos\inst{\ref{CEFCA}} \and S.\,Daflon\inst{\ref{ON}} \and R.\,Dupke\inst{\ref{ON}} \and A.\,Ederoclite\inst{\ref{CEFCA},\ref{CEFCA-UA}} \and C.\,Hernández-Monteagudo\inst{\ref{IAC},\ref{ULL}} \and J.\,Liu\inst{\ref{NAOC}} \and A.\,Marín-Franch\inst{\ref{CEFCA},\ref{CEFCA-UA}} \and C.\,Mendes\,de\,Oliveira\inst{\ref{USP}} \and M.\,Moles\inst{\ref{CEFCA}} \and F.\,Roig\inst{\ref{ON}} \and L.\,Sodré\,Jr.\inst{\ref{USP}} \and K.\,Taylor\inst{\ref{I4}} \and J.\,Varela\inst{\ref{CEFCA}} \and H.\,Vázquez\,Ramió\inst{\ref{CEFCA},\ref{CEFCA-UA}} \and J.\,Zaragoza-Cardiel\inst{\ref{CEFCA}}}
   \institute{
   Centro de Estudios de F\'isica del Cosmos de Arag\'on (CEFCA), Plaza San Juan 1, 44001 Teruel, Spain \label{CEFCA}
   \and
   Unidad Asociada CEFCA--IAA, CEFCA, Unidad Asociada al CSIC por el IAA, Plaza San Juan 1, 44001 Teruel, Spain \label{CEFCA-UA}
   \and
   Instituto de Astrof\'isica de Andaluc\'ia (IAA--CSIC), Glorieta de la Astronom\'ia s/n, 18008 Granada, Spain \label{IAA}
   \and
   Institute of Science and Technology Austria (ISTA), Am Campus 1, A--3400 Klosterneuburg, Austria \label{ISTA}
   \and
   INAF--Osservatorio di Astrofisica e Scienza dello Spazio di Bologna, via Gobetti 93/3, 40129, Bologna, Italy\label{INAF-OAB}
   \and
   Department of Astrophysics, University of Vienna, Türkenschanzstraße 17, 1180 Vienna, Austria \label{UNIVIE}
   \and
   Departamento de Astronomia, Instituto de Astronomia, Geofísica e Ciências Atmosféricas, Universidade de São Paulo, São Paulo, Brazil \label{USP}
   \and
   Observatório Nacional, Rua General José Cristino, 77, São Cristóvão, 20921-400, Rio de Janeiro, RJ, Brazil \label{ON}
   \and
   Independent Researcher \label{IR}
   \and
   Donostia International Physics Center (DIPC), Manuel Lardizabal Ibilbidea, 4, San Sebastián, Spain \label{DIPC}
   % \and
   % Instituto de Física, Universidade Federal da Bahia, 40210-340, Salvador, BA, Brazil \label{UFB}
   \and
   Instituto de Astrofísica de Canarias, C/ Vía Láctea, s/n, E--38205, La Laguna, Tenerife, Spain \label{IAC}
   \and
   Universidad de La Laguna, Avda Francisco Sánchez, E--38206, San Cristóbal de La Laguna, Tenerife, Spain \label{ULL}
   \and
   National Astronomical Observatory of China,  Chinese Academy of Sciences, Beijing, China \label{NAOC}
   \and
   Instruments4, 4121 Pembury Place, La Canada Flintridge, CA 91011, U.S.A. \label{I4}
   }
   \date{\today}

%%%%%%%%%%%%%%%%%%%%%%%%%%%%%%%%%%%%%%%%%%%%%%%%%%%%%%%%%%%%%%
\abstract{Emission lines in galaxy spectra provide unique information on the physical conditions of the interstellar medium, enabling the measurement of fundamental properties such as star formation rate, gas excitation, metallicity, or active galactic nucleus (AGN) activity. In this study, we present the value-added catalogue JOLINES (J-PAS optical line intensities for nebular emission galaxies), which provides emission-line fluxes in galaxies at from the spectrophotometric catalogues of miniJPAS, J-NEP and the J-PAS early data release (EDR). This catalogue will be updated with future data releases, offering a growing resource for the study of emission-line galaxies. To obtain reliable emission-line fluxes from narrow-band photometry, we employed spectral energy distribution (SED) fitting using CIGALE, a robust tool that reconstructs the continuum emission and ensures accurate flux measurements. This method effectively mitigates uncertainties associated with direct continuum subtraction techniques, and systematics such as absorption components in the emission lines. We validate our approach using simulated observations of galaxy spectra with added noise, testing the method's performance across different equivalent width (EW) regimes and emission-line strengths. Additionally, we compare the recovered emission-line fluxes with spectroscopic measurements from the Sloan Digital Sky Survey (SDSS) and the Dark Energy Spectroscopic Instrument (DESI). Our results show a tight correlation between photometric and spectroscopic fluxes, particularly for bright emission lines, with a typical dispersion of $\sim$0.3\,dex. Reliable fluxes are obtained for emission lines with EW $\gtrsim$20\,\AA, in agreement with previous empirical studies. The current catalogue comprises approximately 13\,900 sources with reliable flux measurements in the H$\alpha$+[\ion{N}{ii}] complex and 7\,200 in [\ion{O}{iii}]$\lambda5007$, ensuring statistically robust samples for the brightest optical emission lines. This resource will be expanded in future J-PAS releases, facilitating large-scale studies of star formation, AGN activity, and galaxy evolution.}


%%%%%%%%%%%%%%%%%%%%%%%%%%%%%%%%%%%%%%%%%%%%%%%%%%%%%%%%%%%%%%
\maketitle

\section{Introduction}\label{intro}

%\juan{Everything is wrong, but I can make things worse.}

%The emission lines detected in galaxy spectra provide unique information on the physical properties of the interstellar medium and allow us to measure fundamental properties for studying galaxy evolution, such as star formation rate (SFR), extinction, nebular gas excitation, AGN fraction, etc. Galaxy samples drawn from narrow-band surveys are less affected by selection biases compared to broad-band or spectroscopic surveys, combining a large sky coverage with a deep sensitivity to line emission. However, measuring emission-line fluxes from narrow-band photometry is not straightforward.

The emission lines detected in galaxy spectra provide unique information on the physical properties of the interstellar medium (ISM) and allow us to measure fundamental properties for studying galaxy evolution, such as star formation rate (SFR), extinction, nebular gas excitation, and AGN fraction \citep{kewley19}. Accurate emission-line flux measurements are essential, as they directly trace the physical and chemical conditions in galaxies \citep{osterbrock06} and reveal key processes such as star formation and AGN activity.

To reliably measure emission-line fluxes from narrow-band filters, it is crucial to have an accurate determination of the adjacent continuum emission. A poor assignment of the continuum can significantly affect the measurement of the emission-line flux. Direct subtraction of the continuum emission measured from one or a few filters adjacent to the one containing the emission line is a straightforward method, requiring only a limited number of additional observations. This technique is fast and provides a simple way to estimate line fluxes, particularly when the emission line is strong (e.g. \citealt{ly07,cook19,salzer23}; see \citealt{iglesias-paramo22} for extreme-emission-line galaxies in miniJPAS, where a larger number of continuum filters can be used). However, as the line flux decreases, errors in the determination of the adjacent continuum become more significant, necessitating the use of more sophisticated methods.

Machine learning techniques have been proposed as an alternative, where spectroscopic observations or synthetic models are used to train algorithms to predict the continuum and measure emission or absorption line fluxes in galaxy spectra \citep{ucci18,rhea21,dawson21,jalan24} and in photospectra from narrow-band photometry \citep{martinez-solaeche21}. The advantage of these techniques is their ability to measure emission lines with very low equivalent widths (EWs), down to just a few angstroms, and separate emission-line blends (e.g. H$\alpha$+[\ion{N}{ii}]$\lambda 6548,6583$) based on the properties of the galaxy training sample \citep{martinez-solaeche22}. However, these methods carry a risk: the trained network may propagate biases from the training dataset by overfitting results derived from low signal-to-noise (S/N) data, requiring careful monitoring and control \citep{leung19,dhar22,huertas-company23}.

Another commonly used method is spectral energy distribution (SED) fitting, which reconstructs the continuum by fitting models to the galaxy multi-band spectrum \citep[e.g.][]{lee12,nakajima12}. Although SED fitting has the disadvantage of being influenced by the models used to approximate the continuum, this issue can be mitigated by employing diverse model libraries. By exploring a wide parameter space, SED fitting ensures robust continuum reconstruction and minimises the introduction of artificial features, which are easier to identify and avoid compared to machine-learning methods. Additionally, restricting the use of models that could add unwanted features in the continuum further improves its reliability. In particular, the large number of filters and the wide wavelength coverage of J-PAS surveys offer excellent opportunities for applying SED fitting techniques \citep{mejia-narvaez17,gonzalez-delgado21}. The reliability of this method has been demonstrated by \cite{vilella-rojo15}, \citet{logrono-garcia19}, \citet{lumbreras-calle22}, \citet{breda24} and \citet{rahna25}, by comparing spectroscopic fluxes from SDSS with emission-line fluxes measured from narrow-band photometry in J-PLUS \citep{cenarro19} and miniJPAS \citep{bonoli21}.

In this work we present a catalogue of emission-line fluxes for galaxies in J-PAS, JOLINES (J-\textsc{pas} Optical Line Intensities for Nebular Emission galaxieS). The line fluxes are based on spectrophotometric data extracted from the precursors of the Javalambre-Physics of the Accelerated \textsc{u}niverse Astrophysical Survey (J-PAS; \citealt{benitez14}), i.e. miniJPAS and J-NEP \citep{hernan-caballero23}, and the J-PAS early data release (EDR; Vázquez Ramió et al. in prep.). J-PAS is an unprecedented photometric sky survey of 8500\,deg$^2$, conducted from the Javalambre observatory in Teruel (Spain), using a set of 57 narrow- and several SDSS broad-band filters. Throughout this study, we adopted a flat-$\Lambda$CDM cosmology with $H_0 = 73\, \rm{km\,s^{-1}\,Mpc^{-1}}$ and $\Omega_{\rm m} = 0.27$, $\Omega_\Lambda = 0.73$. In Section\,\ref{obs}, we explain the data obtained from J-PAS surveys. Section\,\ref{results} describes the methodology for continuum fitting and emission-line flux measurements. In Section\,\ref{discuss}, we present the comparison with measured line fluxes for objects with available spectroscopic observations. Finally, in Section\,\ref{summary}, we summarise the results and discuss their implications for future studies.

%%%%%%%%%%%%%%%%%%%%%%%%%%%%%%%%%%%%%%%%%%%%%%%%%%%%%%%%%%%%%%
\begin{figure}[t]
  \centering
  \includegraphics[width = 0.5\textwidth]{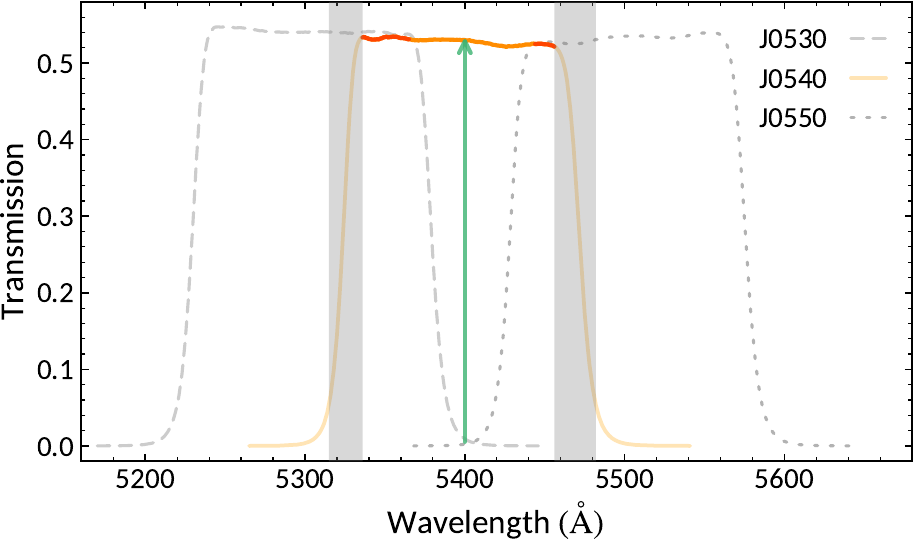}
  \caption{Transmission profile of three J-PAS filters: J0530 (dashed grey line), J0540 (solid yellow line), and J0550 (dotted grey line). A flux measurement for a certain emission line (green arrow) from Table\,\ref{tab_lines} is provided when the observed wavelength falls on the top 5\% maximum transmission of the square filters (orange and red solid line), whereas filters with lines falling at the transmission profile wings ($> 5\%$; shaded grey area) are flagged as contaminated and discarded for the continuum fitting. Note that those cases will have a measurement in one of the adjacent filters, which always overlap at the maximum transmission (in red).}\label{fig_trans}
\end{figure}

\begin{table}[ht!]
\caption{Optical bright emission lines considered to provide measurements derived from narrow-band photometry.}\label{tab_lines}
\centering
\begin{tabular}{cc}
  \bf Name & \bf Wavelength (\AA) \\
  \hline \\[-0.3cm]
  {[\ion{O}{ii}]$\lambda3727$}   & 3726.0, 3728.8 \\
  {[\ion{Ne}{iii}]$\lambda3869$} & 3868.8 \\
  {H$\gamma$+[\ion{O}{iii}]$\lambda4363$}  & 4340.5, 4363.2 \\
  H$\beta$         & 4861.3 \\
  {[\ion{O}{iii}]$\lambda4959$}  & 4958.9 \\
  {[\ion{O}{iii}]$\lambda5007$}  & 5006.8 \\
  {[\ion{O}{i}]$\lambda6300$}    & 6300.3 \\
  {[\ion{N}{ii}]$\lambda6548$}   & 6548.1 \\
  H$\alpha$        & 6562.8 \\
  {[\ion{N}{ii}]$\lambda6583$}   & 6583.5 \\
  {[\ion{S}{ii}]$\lambda6724$}   & 6716.4, 6730.8 \\
  {[\ion{S}{iii}]$\lambda9069$}  & 9068.6 \\
  {[\ion{S}{iii}]$\lambda9531$}  & 9531.1 \\
  \hline
\end{tabular}
\end{table}

\begin{table*}[t!]
\caption{Parameters and their respective sampling ranges for the various modules used in CIGALE v2025.0 SED fitting.} 
\centering
\setlength{\tabcolsep}{1.mm}
\begin{tabular}{cc}
\bf Parameters &  \bf Sampling range\\
\hline\\[-1.5ex]
\multicolumn{2}{c}{\it Star formation history: delayed model} \\[0.5ex]
Age of the main population &  5, 8, 10, 13 Gyr  \\
e-folding time & 1, 5 Gyr \\ 
Age of the young burst &  1, 5, 10, 100, 500 Myr  \\
e-folding time young burst & 5, 50, 100 Myr \\ 
Burst stellar mass fraction & 0.0, 0.2, 0.4, 0.6, 0.8, 0.99\\[0.5ex]
\hline\\[-1.5ex]
\multicolumn{2}{c}{\it Simple Stellar population: Charlot \& Bruzual (2019)} \\[0.5ex] %\url{http://www.bruzual.org/CB19} \citet{bruzual03}} 
Initial Mass Function & \citet{chabrier03} \\
Metallicity & 0.004, 0.01, 0.02, 0.03 \\
Upper IMF limit & 100 M$_\odot$ \\[0.5ex]
\hline\\[-1.5ex]
\multicolumn{2}{c}{\it Nebular emission} \\[0.5ex]
Ionisation parameter (log\,U) &  -2.0, -3.0, -4.0 \\
Gas metallicity ($Z_\text{gas}$) & 0.004, 0.011, 0.020, 0.033 \\
Electron density ($\rm n_e$)     & 100 cm$^{-3}$ \\[0.5ex]
\hline\\[-1.5ex]
\multicolumn{2}{c}{\it Dust extinction} \\[0.5ex]
Dust attenuation law & modified \citet{charlot00}\\
ISM \textit{V}-band attenuation $\left(A^\text{ISM}_\text{V} \right)$ & 0.0, 0.2, 0.4, 0.6, 0.8, 1.0, 1.5, 2.0, 2.5, 3.0 \\
Ratio of ISM to total attenuation $\left(\mu = \frac{A^\text{ISM}_\text{V}}{A^\text{ISM}_\text{V} + A^\text{BC}_\text{V}} \right)$ & 0.44 \\
Power law slope of ISM attenuation & -0.7 \\
Power law slope of birth clouds attenuation & -1.3 \\[0.5ex]
\hline\\[-1.5ex]
Number of models per redshift & 345\,600 \\[0.5ex]
\hline
\label{tab_cigpars}
\end{tabular}
\end{table*}

\section{Observations and methods}\label{obs}

The sample of galaxies in JOLINES was drawn from a query to the miniJPAS, J-NEP, and EDR dual catalogues\footnote{\textit{r}-band detection in miniJPAS and J-NEP, \textit{i}-band detection in EDR.} by selecting targets with a low probability of being star-like (i.e. stars or quasars; \texttt{CLASS\_STAR} < 0.1) and with no flags in their photometric measurements (i.e. all \texttt{FLAGS} and \texttt{MASK\_FLAGS} set to 0). An additional cut was applied to select sources with \textit{i}-band magnitudes brighter than $21\, \rm{mag}$, in order to exclude objects with very low S/N in the continuum determinations, and with photo-$z$ $< 0.4$, to ensure that the H$\alpha$ line remains within the rest-frame spectral range. To derive emission-line fluxes from the photometric measurements, we performed SED fitting with CIGALE\footnote{\url{https://cigale.lam.fr}} (Code Investigating GALaxy Emission; \citealt{boquien19}), owing to its versatility and efficiency in computing best-fits for large galaxy samples.

\subsection{Flux catalogues and photo-zs}\label{subsec_fluxes}
Fluxes from the miniJPAS, J-NEP, and EDR catalogues were extracted from the \texttt{FLUX\_APER\_COR\_3\_0} column in the dual catalogue, which represents photometric fluxes measured within a three arcsecond aperture and corrected for aperture effects for sources detected in the \textit{r} (miniJPAS and J-NEP) or the \textit{i} bands (EDR). The aperture correction assumes point-like sources in the dual catalogue, ensuring consistent flux measurements despite challenges in extended sources like galaxies. We selected the \texttt{APER\_COR\_3\_0} aperture specifically to ensure robust spectrophotometric calibration across the full filter set. While this fixed aperture may not capture the total flux of extended sources, it is essential for minimising systematic scatter among continuum filters caused by varying observing conditions, such as changes in seeing and background depth (Vázquez Ramió et al. in prep.). To address potential flux losses in extended sources, the photospectrum of each galaxy was scaled by the factor required to match the \textit{r}-band flux in \texttt{FLUX\_APER\_COR\_3\_0} with that in \texttt{FLUX\_AUTO} measurements. Photometric redshifts for individual sources are available in all catalogues and were calculated using the method described by \citet{hernan-caballero21}. This approach employs a customised version of \textsc{Lephare} \citep{arnouts11}, optimised for J-PAS observations with a selection of 50 synthetic galaxy templates from CIGALE to enhance photo-z accuracy for a subsample of galaxies. Additionally, a ``galaxy locus recalibration'' was performed, which adjusts systematic photometric offsets via median galaxy colours \citep{hernan-caballero23}. Finally, redshift probability distributions were combined through conflation\footnote{Conflation refers to the normalised product of independent probability density functions of the same underlying variable. This operation produces a combined estimate with reduced uncertainty.}, leveraging the precision of miniJPAS narrow-band filters and the depth of broad-band imaging from the Hyper-SuprimeCam Subaru Strategic Program \citep{hernan-caballero24}. This integration improved the miniJPAS photo-$z$ estimates, particularly for faint sources. Overall, the majority of sources in miniJPAS, J-NEP, and EDR catalogues with $i < 21.5\, \rm{mag}$ have photo-z uncertainties of $|\Delta z| < 0.3\,\%$ \citep{hernan-caballero23,hernan-caballero24}.

The selection of emission-line filters was a key step in our methodology. We identified filters whose transmission profiles may include bright emission lines (see Table\,\ref{tab_lines}). Figure\,\ref{fig_trans} illustrates the transmission profiles of three J-PAS narrow-band filters, represented as dashed, solid, and dotted lines. A filter is considered to contain an emission line when the observed wavelength of the line falls within the maximum transmission in the filter's square profile. For instance, the maximum transmission region of the J0540 filter is highlighted in orange in Fig.\,\ref{fig_trans}, while the transmission wings are shown in yellow. A flux measurement is provided for a certain emission line (green arrow) when the observed line wavelength falls within this flat maximum transmission region. Conversely, potential contamination from nearby emission lines can occur when a transition falls within the filter's transmission wings, due to uncertainties in the contribution of the line to the integrated flux. Thus, measurements affected by potential contamination in the transmission wings, at transmission levels exceeding 5\% (shaded grey area in Fig.\,\ref{fig_trans}), are flagged and excluded from the continuum fitting. Nevertheless, a flux measurement in those cases will be provided by adjacent filters, which always overlap at the maximum transmission (red solid line). We considered a filter as suitable continuum for SED fitting only if it contained no bright emission lines within either the maximum transmission region or the transmission wings.

Emission lines falling within $\sim$150\,\AA\ in observed wavelength --\,comparable to the typical width of the J-PAS narrow-band filters\,-- may be blended depending on the galaxy redshift. This situation frequently arises for pairs such as H$\beta$ and [\ion{O}{iii}]$\lambda4959$, the [\ion{O}{iii}]$\lambda4959,5007$ doublet, or H$\alpha$ and the [\ion{N}{ii}]$\lambda6548,6583$ doublet. In these cases, the JOLINES catalogue provides the total blended flux, and individual line intensities are recovered when possible using flux measurements from adjacent overlapping filters and the theoretical nebular ratios for the [\ion{O}{iii}] and [\ion{N}{ii}] doublets (see discussion in Sections\,\ref{sec_desi} and \ref{sec_spec}). Conversely, doublets with small wavelength separations, such as [\ion{O}{ii}]$\lambda3726,3729$ and [\ion{S}{ii}]$\lambda6716,6731$, cannot be deblended in most cases, since their intrinsic ratios depend strongly on electron density and cannot be disentangled on theoretical grounds.

\subsection{Continuum subtraction}\label{subsec_cont}
For each galaxy, the continuum emission was estimated by fitting the photospectrum after masking the filters containing the bright emission lines listed in Table\,\ref{tab_lines}. CIGALE v2025.0 was chosen for the SED fitting procedure due to its efficiency in processing large samples of galaxies and its capability to provide Bayesian fitting estimates. Initially, spectrophotometric fluxes from the J-PAS catalogues were converted to mJy and transformed to the CIGALE input table format. Filter transmission profiles for J-PAS filters are available at the Spanish Virtual Observatory (SVO) service\footnote{\url{https://svo.cab.inta-csic.es}} \citep{rodrigo12,rodrigo20}. Since the main goal of the SED fitting process is to perform an accurate determination of the continuum level, we only take into account continuum-dominated filters to avoid possible biases in the best-fit continuum. These could be caused, for example, by the presence of very bright emission lines that may bias the best-fit solutions towards bluer continua, required to reproduce the line properties. On the other hand, the nebular continuum was taken into account for the fitting, since this component may contribute significantly to the optical/UV range in low-mass star-forming galaxies \citep[e.g.][]{reines10,cardoso22,lumbreras-calle22} and radiogalaxies with strong emission lines \citep{dickson95,solorzano-inarrea04,holt07}. The parameters of the nebular module in CIGALE were restricted to a relatively narrow grid, with three ionisation parameter values, four gas metallicities, and a single electron density of $100\,\mathrm{cm^{-3}}$. This choice was motivated by the fact that emission lines are not included in the fitting process, so a compact parameter grid was adopted to reduce the computation time required for the main model library. Line flux uncertainties are derived from the quadratic propagation of the uncertainties associated with the photometric measurements and the best-fit continuum estimate.

To accommodate a wide diversity of galaxy properties in the continuum subtraction process, we used models encompassing a broad range of parameters, including stellar ages, star formation rates (SFRs), star formation histories, dust extinction, and metallicity. Table\,\ref{tab_cigpars} shows a summary of the fitting modules adopted for CIGALE, the most relevant parameters, and their respective sampling ranges. This comprehensive grid, comprising approximately $\sim$14 million models ($345\,600$ per redshift in $\sim 40$ redshift bins between $z = 0$ and $0.4$), enabled accurate reproduction of the continuum emission in J-PAS galaxies.
\begin{figure}[t]
  \centering
  \includegraphics[width = 0.5\textwidth]{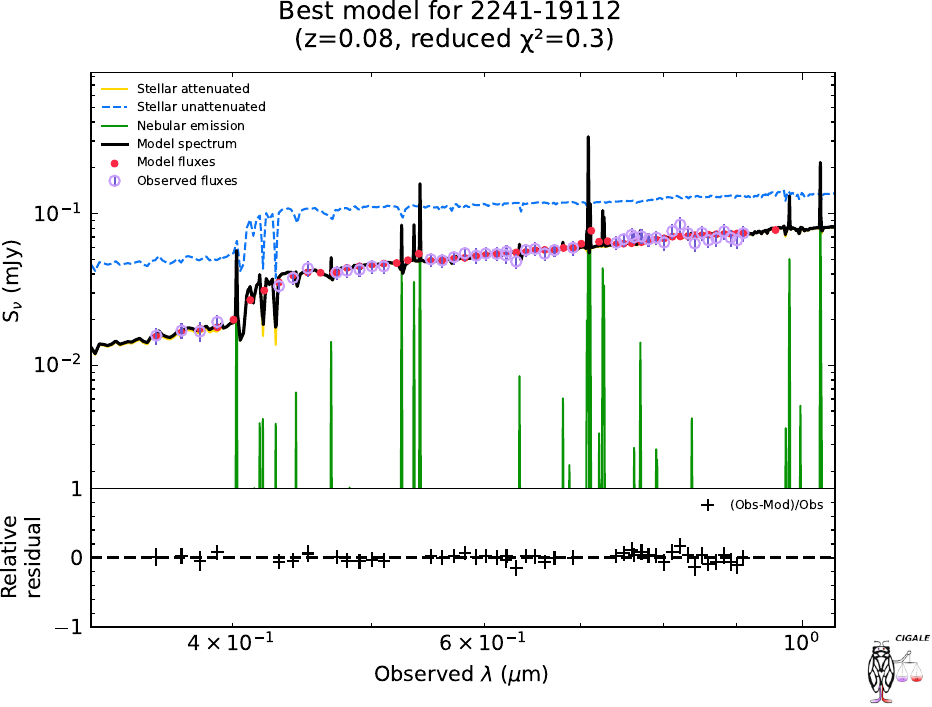}
  \caption{Best-fit SED obtained with CIGALE for the miniJPAS galaxy 2241--19112. The upper panel displays the observed fluxes (purple circles), the best-fit fluxes derived from the fitting (red dots), the stellar continuum before (dashed blue line) and after (solid yellow line) applying dust attenuation, the nebular line plus continuum emission (solid green line), and the total model (solid black line). The lower panel presents the relative differences between the observed and modelled fluxes.}\label{fig_cigale}
\end{figure}

\begin{figure*}[t]
  \centering
  \includegraphics[width = 0.5\textwidth]{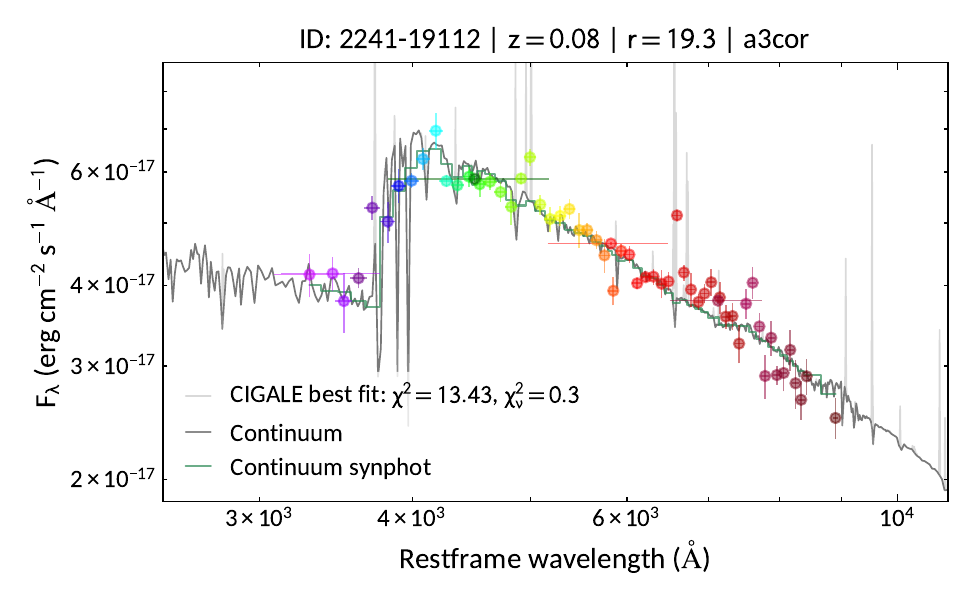}~
  \includegraphics[width = 0.5\textwidth]{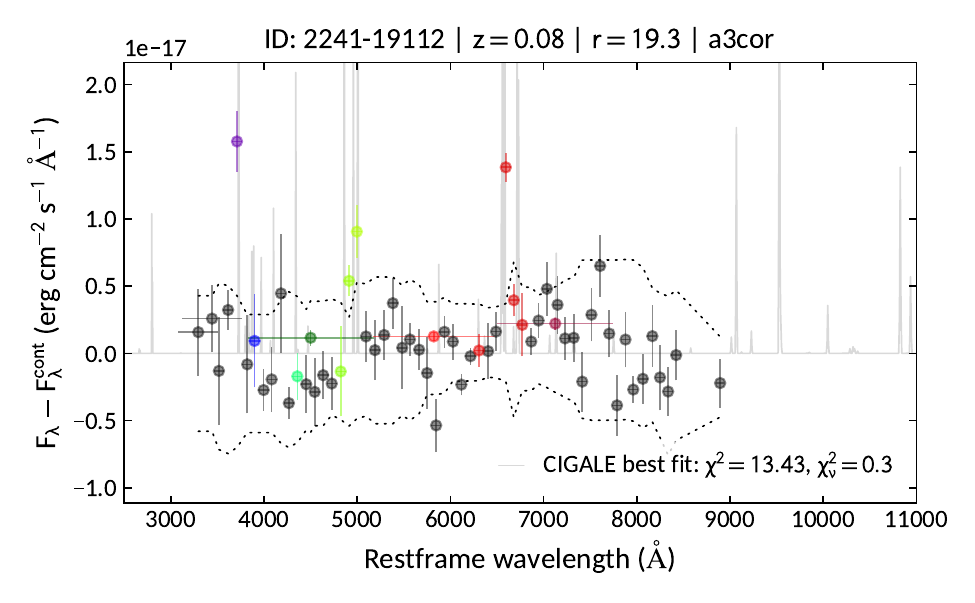}
  \caption{Extraction of emission-line fluxes from the photospectrum of the galaxy 2241--19112, using the CIGALE best-fit model for the continuum subtraction. \textit{Left:} the observed photospectrum (coloured dots) is compared with the best-fit model (solid light-grey line) and its corresponding continuum (solid dark grey line). The solid green step line indicates the synthetic photometry for the best-fit continuum distribution obtained for the J-PAS filters. \textit{Right:} dots show the continuum-subtracted photospectrum, in colour for filters including lines in Table\,\ref{tab_lines}, in grey for continuum filters used in the SED fitting. Dotted black lines indicate the $3\sigma$ uncertainty, derived from the scatter in the residual values of the continuum-subtracted photospectrum. The solid light-grey line shows the emission-line spectrum for the CIGALE best-fit model.}\label{fig_contsub}
\end{figure*}

Figure\,\ref{fig_cigale} shows an example of the best-fit model obtained for the miniJPAS galaxy \textsc{tile-id--number} = 2241--19112. In this example, the best-fit attenuated stellar emission provides a robust description of the continuum, allowing us to subtract this component from the observed photospectrum. This is shown in more detail in Fig.\,\ref{fig_contsub} (left panel), where the modelled continuum (in dark grey), accounts for the absorption lines due to the old stellar populations, allowing us to correct the line fluxes measured in the narrow-bands (colourful dots). Bright emission lines of [\ion{O}{ii}]$\lambda3727$, H$\beta$ + [\ion{O}{iii}]$\lambda4959$, [\ion{O}{iii}]$\lambda5007$, and H$\alpha$ + [\ion{N}{ii}]$\lambda6548$ stand out in the continuum-subtracted photospectrum shown in Fig.\,\ref{fig_cigale} (right panel). The scatter in the continuum fluxes (dotted black lines) is mainly caused by differences in the observing conditions, since all filters are acquired on different nights, typically over a period of a few months. Figure\,\ref{fig_contsub} shows how continuum subtraction using only line-adjacent filters could limit our measurements to brighter lines, due to uncertainties in the continuum level.

%Overlap in the peak transmission regions of adjacent filters allows for duplicate measurements of emission lines whose observed wavelengths fall within the overlapping range.
In some cases, an emission line may fall within the overlap region of two adjacent filters (see Fig.\,\ref{fig_trans}), allowing it to be detected in both filters independently. Duplicated measurements in the catalogue were averaged to improve the S/N, ensuring more robust flux estimates. For filters containing two bright emission lines, individual line fluxes were determined in cases where one of the lines was also detected in the overlap region of an adjacent filter. In such cases, the flux measurement from the overlapping filter was used to disentangle the contributions of the two lines, and the associated uncertainties in the flux measurements were propagated accordingly. Additionally, theoretical line ratios were employed to resolve degeneracies in cases where filters contained multiple emission lines that could not be disentangled solely through measurements. Specifically, we used the theoretical ratios of [\ion{O}{iii}]$\lambda5007$/[\ion{O}{iii}]$\lambda4959$ $= 2.98$ and [\ion{N}{ii}]$\lambda6583$/[\ion{N}{ii}]$\lambda6548$ $= 2.94$ provided by \textsc{PyNeb} \citep{luridiana15} to estimate the individual line fluxes, corresponding to an electron temperature and density of $T_\text{e} = 10^{4}\, \rm{K}$ and $n_\text{e} = 100\, \rm{cm^{-3}}$, respectively. For instance, flux measurements involving H$\beta$ + [\ion{O}{iii}]$\lambda4959$, H$\alpha$ + [\ion{N}{ii}]$\lambda6548$, and H$\alpha$ + [\ion{N}{ii}]$\lambda6583$ were decomposed into their individual line intensities using these known ratios.

%%%%%%%%%%%%%%%%%%%%%%%%%%%%%%%%%%%%%%%%%%%%%%%%%%%%%%%%%%%%%%

\section{Results}\label{results}

The robustness of the method developed to measure emission-line fluxes (Section\,\ref{obs}) was tested using simulated observations based on galaxy models with added noise (Section\,\ref{sec_mock}), synthetic spectrophotometry derived from the DESI spectroscopic survey (Section\,\ref{sec_desi}), and a comparison between the estimated line fluxes for galaxies in the J-PAS catalogues and available spectroscopic measurements (Section\,\ref{sec_spec}).

\begin{figure*}[t]
  \centering
  \subfigure[]{\includegraphics[width = 0.5\textwidth]{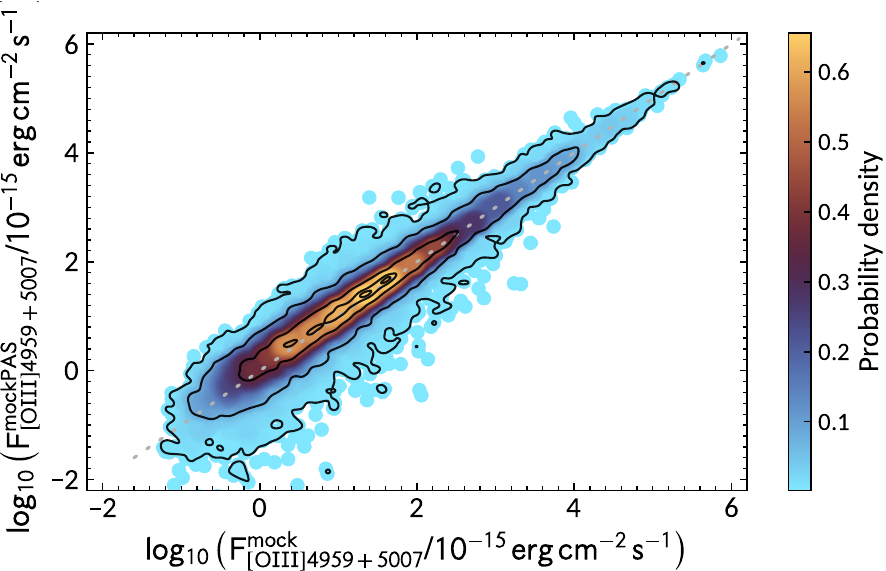}\label{subfig_fo3}}~
  \subfigure[]{\includegraphics[width = 0.5\textwidth]{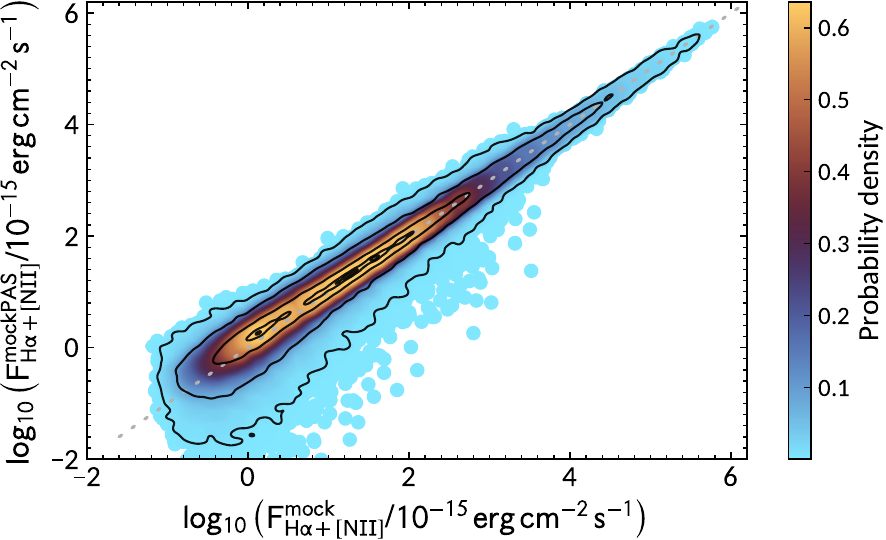}\label{subfig_fha}}
  %\subfigure[]{\includegraphics[width = 0.5\textwidth]{figs/mock6_Ro3.pdf}\label{subfig_ro3_fo3}}~
  %\subfigure[]{\includegraphics[width = 0.5\textwidth]{figs/mock6_RHa.pdf}\label{subfig_rha_fha}}
  \subfigure[]{\includegraphics[width = 0.5\textwidth]{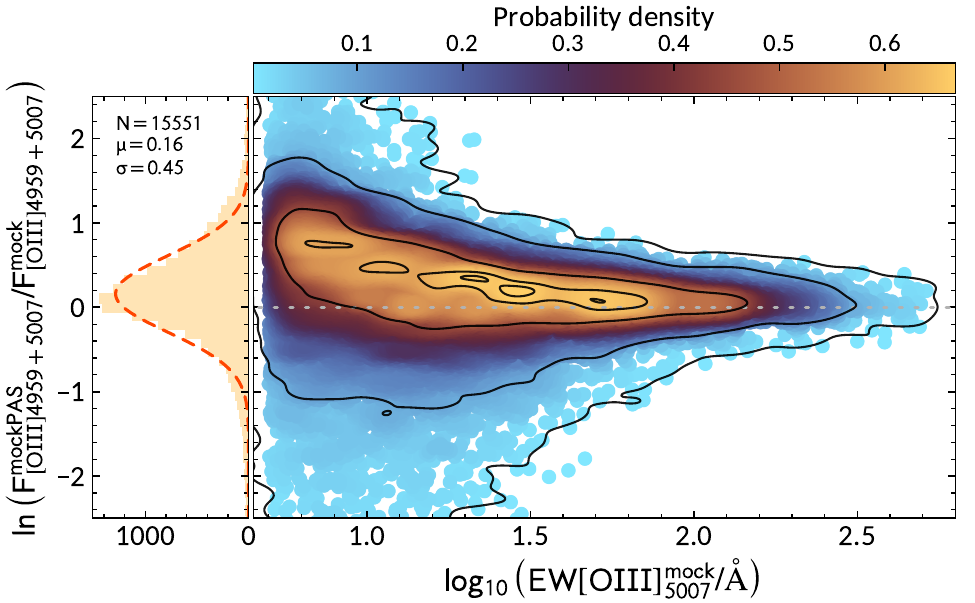}\label{subfig_ro3_ewo3}}~
  \subfigure[]{\includegraphics[width = 0.5\textwidth]{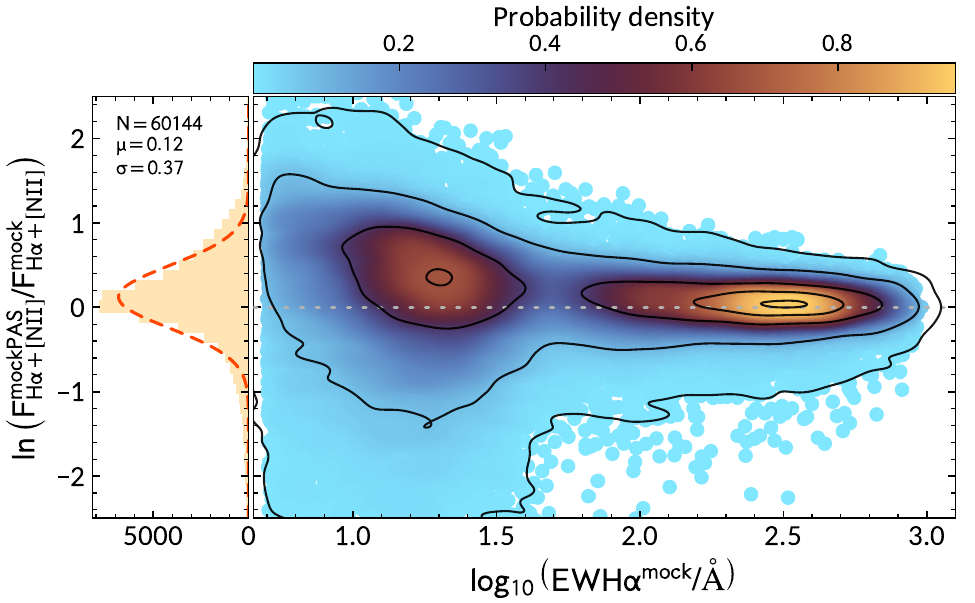}\label{subfig_rha_ewha}}
  \caption{The upper panels show density plots of the emission-line fluxes for [\ion{O}{iii}]$\lambda4959,5007$ (a) and H$\alpha$ + [\ion{N}{ii}]$\lambda6548,6583$ (b) derived from the mock catalogue spectra with the J-PAS filter system, using the SED-fitting-based continuum subtraction method described in Section\,\ref{obs}. The J-PAS fluxes are compared with the input line fluxes in the mock catalogue. The lower panels show the line flux ratios between the J-PAS line fluxes and the mock input fluxes for [\ion{O}{iii}]$\lambda 4959,5007$ (c) and H$\alpha$ + [\ion{N}{ii}]$\lambda 6548,6583$ (d), plotted as a function of the mock equivalent width of [\ion{O}{iii}]$\lambda 5007$ and H$\alpha$, respectively. In dashed grey style, we show the one-to-one lines in the upper panels and the constant unity value in the lower panels. The side vertical panels in (c) and (d) show the marginal distribution (grey histogram) and the corresponding Gaussian fit (dashed red line) of the J-PAS to mock line flux ratios. In all panels, the black contours represent the 1, 10, 50, 90, and 99\% percentiles of the two-dimensional probability distribution.}\label{fig_mock}
\end{figure*}

\subsection{Comparison with simulated galaxies}\label{sec_mock}
A new mock catalogue was generated to simulate J-PAS observations, providing a controlled environment to assess the accuracy of the emission-line flux recovery method and to quantify the uncertainties associated with the SED-fitting procedure itself. Unlike real observations, the use of a mock catalogue allows us to explore a much wider range of physical parameters --\,such as stellar population age, dust attenuation, metallicity, and emission-line strength\,-- than is accessible with current spectroscopic data. This controlled setup enables us to isolate and characterise the effects of model assumptions (e.g. the choice of stellar population synthesis, dust attenuation law, or parameter sampling) on the recovered fluxes, thus providing a reference framework for interpreting the results obtained from real spectra in Section\,\ref{sec_desi}. The models were created with the CIGALE code, sampling the redshift range $0 < z < 0.4$, and adopting a configuration independent from that used for fitting, including different stellar population models, dust extinction laws, and parameter grids (see Table~\ref{tab_mockpars}). Specifically, the mock catalogue employs the \citet{bruzual03} models with a \citet{salpeter55} initial mass function (IMF) and the \citet{calzetti00} dust attenuation law. The parameter values sampled for the mock grid are entirely different from those in the fitting library, ensuring that no model used for fitting appears among the synthetic inputs. This design not only prevents any overlap between the input and fitting models but also allows us to calibrate how robust the method is at reproducing the continuum around emission lines when the true continuum shape lies completely outside the fitting grid. Finally, the larger number of models in the fitting library, relative to the mock grid, was intended to account for potential numerical degeneracies and model biases during the fitting process. Synthetic photometry was then produced from these models using the J-PAS filter set available in the SVO, and Gaussian noise was stochastically added at a 10\% level to reproduce typical observational uncertainties. While a uniform noise level provides a reasonable approximation for most sources, we note that a more detailed noise model --\,accounting for variations in source magnitude, line equivalent width, and observing conditions\,-- would be more realistic for specific cases. However, for the mock catalogue we adopted a simpler 10\% noise level in order to isolate systematics arising solely from the modelling and fitting procedures. The impact of realistic observational noise is taken into account in Sections~\ref{sec_desi} and \ref{sec_spec}, which rely on real DESI and J-PAS observations, respectively.

The synthetic spectra were subsequently processed using the methodology described in Section\,\ref{obs}. Specifically, the data were formatted into input tables for CIGALE, with filters containing emission lines masked, both at the peak of the emission and in the transmission wings, to ensure unbiased continuum fitting. These spectra were then fitted with CIGALE using the parameters detailed in Table\,\ref{tab_cigpars}. Following this, the emission-line fluxes were obtained by subtracting the continuum emission and applying the arithmetic procedures outlined in Section\,\ref{obs} to decompose blended fluxes into individual line fluxes.

In Fig.\,\ref{fig_mock}, we present the recovered line fluxes for the total [\ion{O}{iii}]$\lambda4959,5007$ flux and the H$\alpha$ + [\ion{N}{ii}]$\lambda6548,6583$ blend. These fluxes are shown as a function of the model input fluxes (Figs\,\ref{subfig_fo3} and \ref{subfig_fha}) and the equivalent width values (Figs\,\ref{subfig_ro3_ewo3} and \ref{subfig_rha_ewha}). Bright emission lines with fluxes of $\gtrsim 10^{-14}\, \rm{erg\,s^{-1}\,cm^{-2}}$ or those relatively bright compared to their continua (EW\,$\gtrsim$10\,\AA) are recovered with a dispersion $\lesssim 0.2\, \rm{dex}$ and without significant bias for EW values above $\sim$20\,\AA, as shown in Figs\,\ref{subfig_ro3_ewo3} and \ref{subfig_rha_ewha}. However, at low equivalent widths (EW $\lesssim$ 10\,\AA), a bias of $\sim 0.1$--$0.3$ towards higher recovered fluxes is observed. This is caused by the loss of contrast between the emission line and the continuum, which increases the uncertainty of the line flux measurement in the narrower filters, challenging the accurate recovery of faint lines blended with the continuum.

Figs\,\ref{subfig_ro3_ewo3} and \ref{subfig_rha_ewha} demonstrate that line fluxes with typical uncertainties of $0.3\, \rm{dex}$ are achieved for most synthetic galaxies with emission lines of EW $\gtrsim$20\,\AA. This result aligns with the detectability threshold of EW $\gtrsim$ 25\,\AA, empirically derived by \citet{breda24} for miniJPAS using polynomial plus Gaussian fitting to extract emission-line fluxes from photospectra. Overall, the results confirm the reliability of the developed method for recovering emission-line fluxes, particularly for lines with high S/N and moderate to large equivalent widths.

\begin{figure*}[t]
  \centering
  \subfigure[]{\includegraphics[width = 0.49\textwidth]{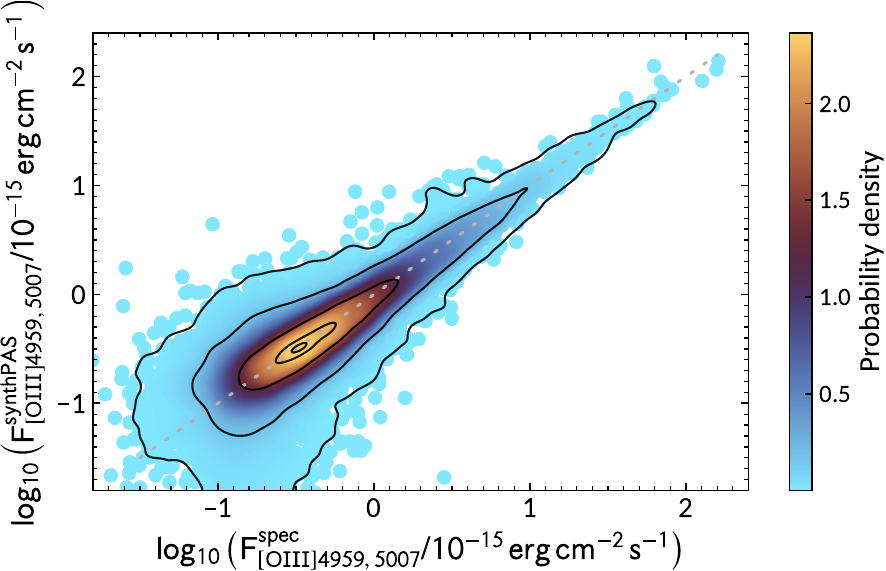}\label{subfig_desi_Fo3}}~
  \subfigure[]{\includegraphics[width = 0.49\textwidth]{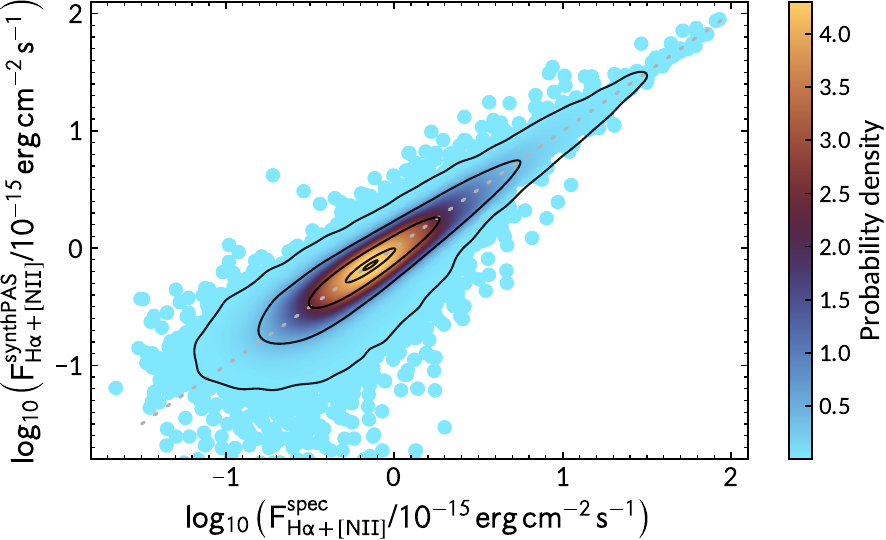}\label{subfig_desi_Fha}}
  \subfigure[]{\includegraphics[width = 0.49\textwidth]{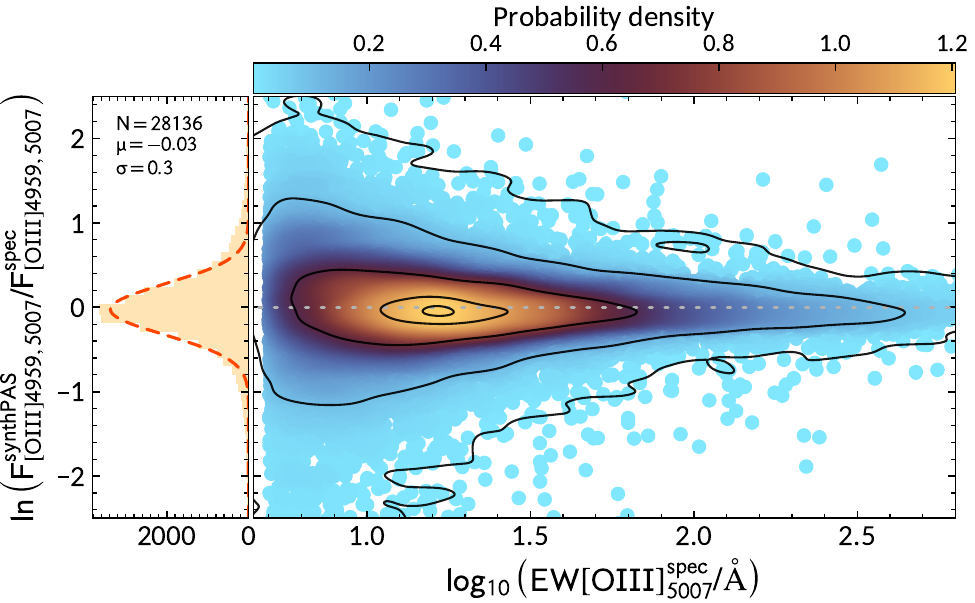}\label{subfig_desi_ewo3}}~
  \subfigure[]{\includegraphics[width = 0.49\textwidth]{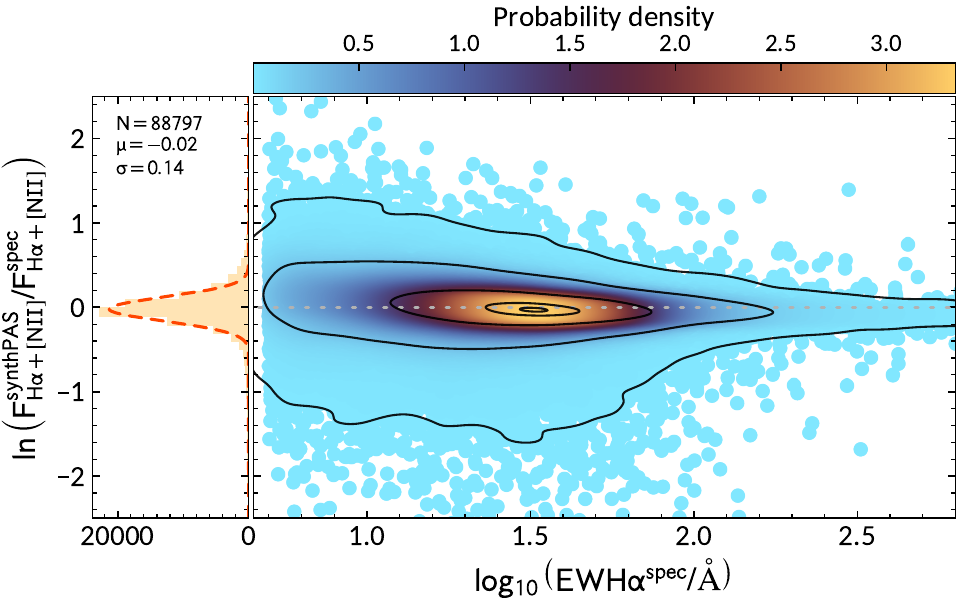}\label{subfig_desi_ewha}}
  \caption{Same as Fig.~\ref{fig_mock}, but comparing the J-PAS fluxes obtained from the synthetic photometry of DESI EDR spectra with the spectroscopic line fluxes from the FastSpecFit value-added catalogue (\citealt{moustakas23}; Moustakas et al. in prep.). Panels (a) and (b) show the density plots for [\ion{O}{iii}]$\lambda4959,5007$ and H$\alpha$+[\ion{N}{ii}]$\lambda6548,6583$, respectively. Panels (c) and (d) display the corresponding J-PAS to spectroscopic flux ratios as a function of the spectroscopic equivalent widths.}\label{fig_desi}
  %The upper panels show the density plots of the emission-line fluxes for [\ion{O}{iii}]$\lambda 4959,5007$ (a) and H$\alpha$ + [\ion{N}{ii}]$\lambda 6548,6583$ (b) derived from the synthetic photometry of DESI EDR \citep{desi24} spectra with the J-PAS filter system, using the SED fitting-based continuum subtraction method described in Section\,\ref{obs}. The J-PAS fluxes are compared with the spectroscopic line fluxes from the FastSpecFit value-added catalogue (\citealt{moustakas23}; Moustakas et al. in prep.). The lower panels show the line flux ratios between J-PAS line fluxes and the spectroscopic measurements for [\ion{O}{iii}]$\lambda 4959,5007$ (c) and H$\alpha$ + [\ion{N}{ii}]$\lambda 6548,6583$ (d), plotted as a function of the spectroscopic equivalent width of [\ion{O}{iii}]$\lambda 5007$ and H$\alpha$, respectively. In dashed grey style, we show the one-to-one lines in the upper panels and the constant unity value in the lower panels. The side vertical panels in (c) and (d) show the marginal distribution (grey histogram) and the corresponding Gaussian fit (dashed red line) of the J-PAS to spectroscopic line flux ratios. In all panels, the black contours represent the 1, 10, 50, 90, and 99\% percentiles of the two-dimensional probability distribution.
\end{figure*}

\begin{figure*}[t]
  \centering
  \subfigure[]{\includegraphics[width = 0.49\textwidth]{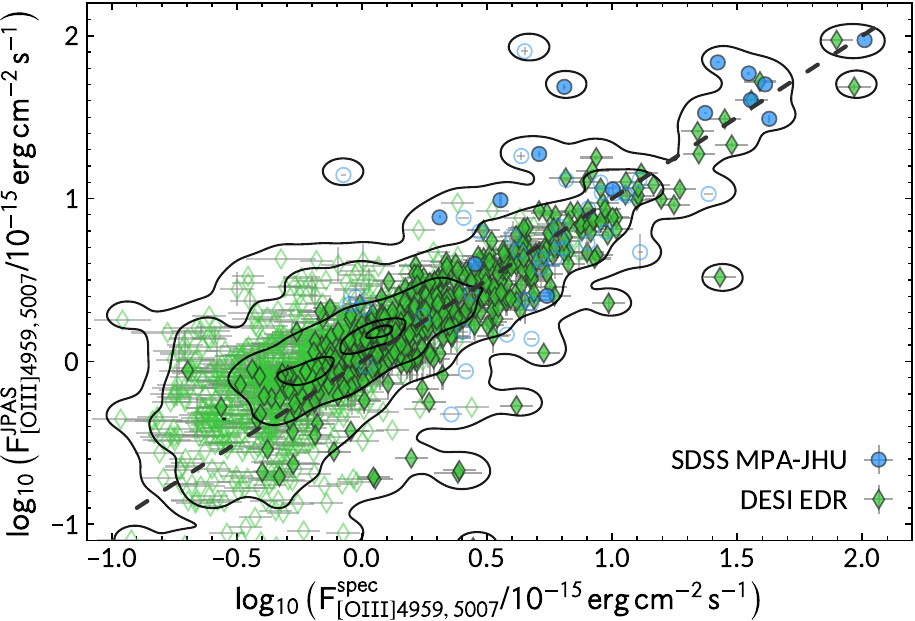}\label{subfig_spec_Fo3}}~
  \subfigure[]{\includegraphics[width = 0.49\textwidth]{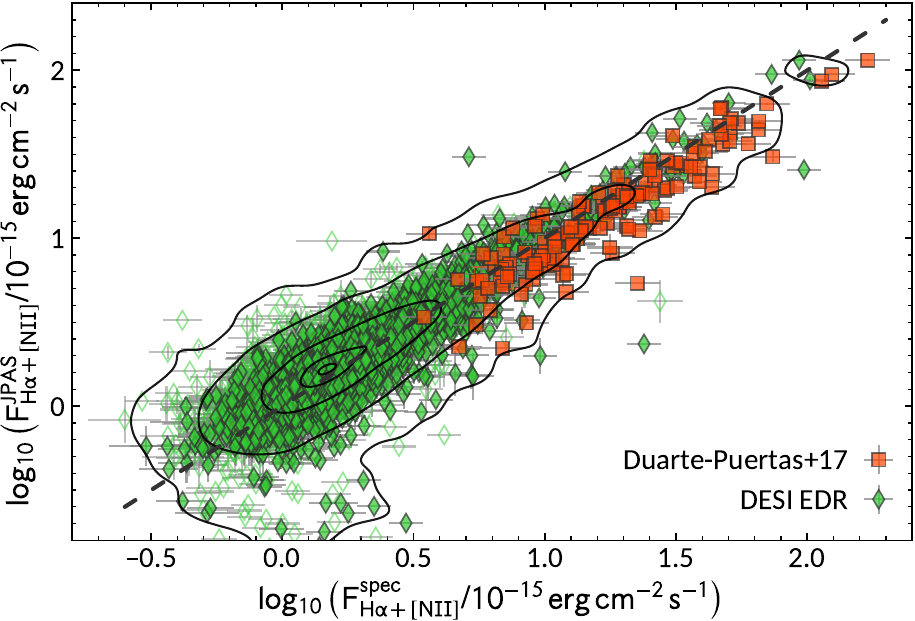}\label{subfig_spec_Fha}}
  \subfigure[]{\includegraphics[width = 0.49\textwidth]{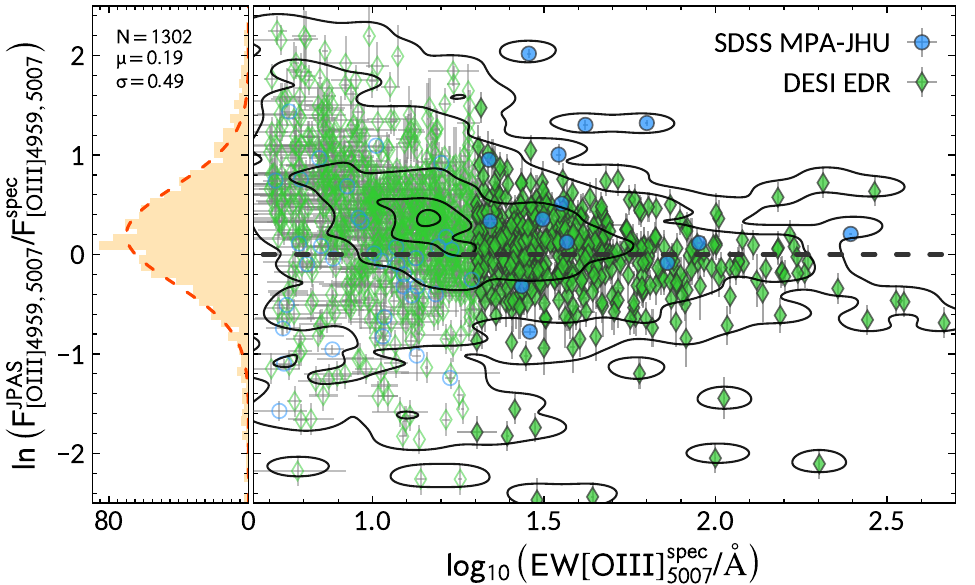}\label{subfig_spec_ewo3}}~
  \subfigure[]{\includegraphics[width = 0.49\textwidth]{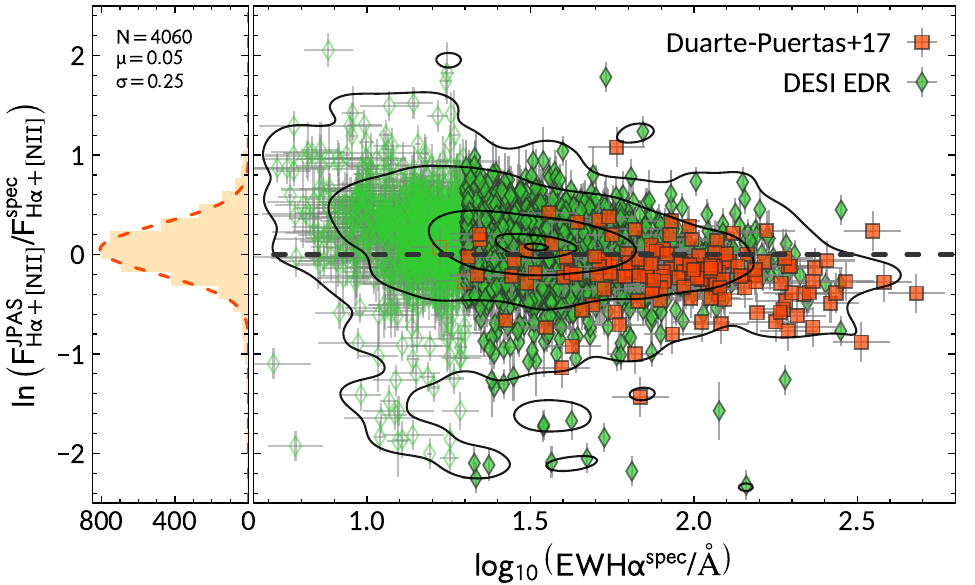}\label{subfig_spec_ewha}}
  \caption{Same as Fig.~\ref{fig_mock}, but using real miniJPAS, J-NEP and J-PAS EDR photospectra. JOLINES fluxes are compared with SDSS spectroscopic measurements from the MPA--JHU DR8 catalogue (blue circles), aperture-corrected H$\alpha$ fluxes from \citet{duartepuertas17} (red squares), and DESI DR1 line fluxes (\citealt{moustakas23}; Moustakas et al. in prep.; green diamonds). Open and filled symbols denoting EW ranges of 5--20\,\AA\ and $>$20\,\AA, respectively. Black contours trace the 1, 10, 50, 90, and 99\% percentiles.}\label{fig_spec}
\end{figure*}

\begin{figure*}[t]
  \centering
  \subfigure[]{\includegraphics[width = 0.33\textwidth]{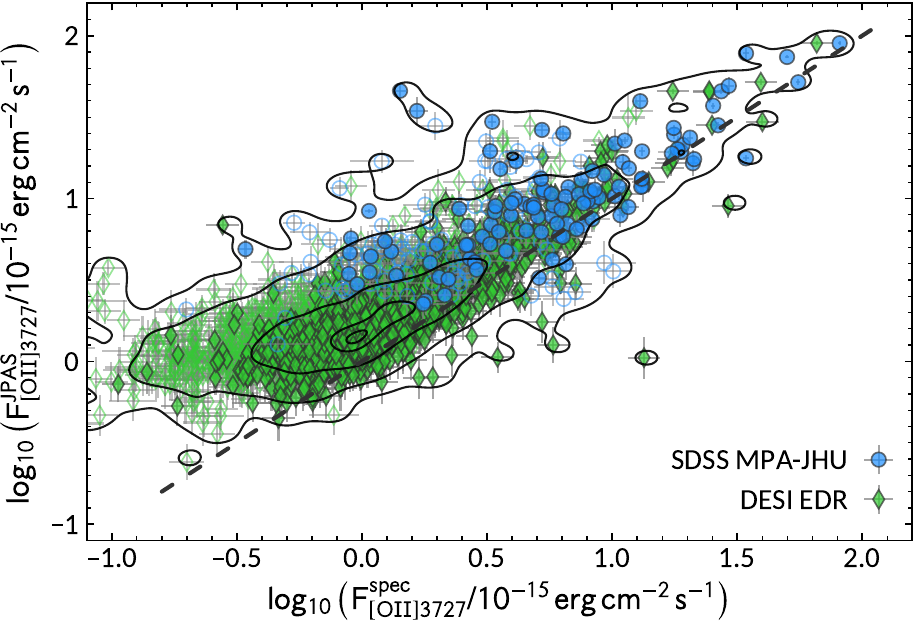}\label{subfig_spec_Fo2}}~
  \subfigure[]{\includegraphics[width = 0.33\textwidth]{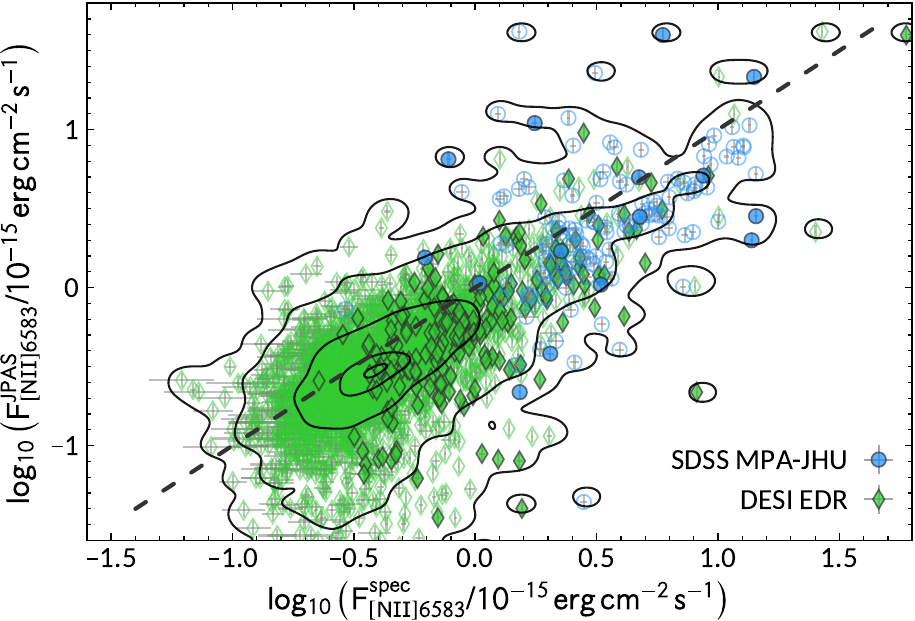}\label{subfig_spec_Fn2}}~
  \subfigure[]{\includegraphics[width = 0.33\textwidth]{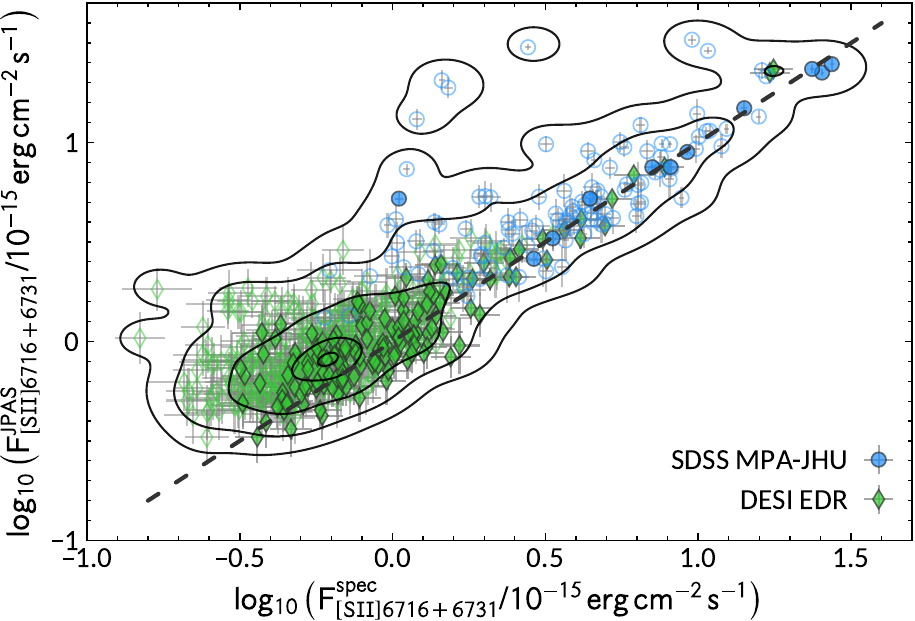}\label{subfig_spec_Fs2}}
  \caption{Emission-line fluxes for [\ion{O}{ii}]$\lambda3726,3729$ (a), [\ion{N}{ii}]$\lambda6583$ (b), and [\ion{S}{ii}]$\lambda6716,6731$ (c) derived from the miniJPAS, J-NEP and J-PAS EDR photospectra. JOLINES fluxes are compared with SDSS spectroscopic measurements from the MPA--JHU DR8 catalogue (blue circles; rescaled using the aperture-corrected H$\alpha$ fluxes from \citealt{duartepuertas17}) and DESI DR1 line fluxes (\citealt{moustakas23}; Moustakas et al. in prep.; green diamonds). For panel (b), a $\sim22\%$ contribution of [\ion{N}{ii}]$\lambda6583$ to the H$\alpha$+[\ion{N}{ii}] blend is assumed. Open and filled symbols denote equivalent widths of 5--20\,\AA\ and $>20$\,\AA, respectively. Dashed grey lines mark the one-to-one relation, and black contours indicate the 1, 10, 50, 90, and 99\% percentiles of the two-dimensional probability distribution.}\label{fig_spec_faint}
  %Emission-line fluxes for [\ion{O}{ii}]$\lambda 3726,3729$ (a), [\ion{N}{ii}]$\lambda 6583$ (b), and [\ion{S}{ii}]$\lambda 6716,6731$ (c) derived from the miniJPAS and J-PAS EDR photospectra using the SED fitting-based continuum subtraction method described in Section\,\ref{obs}. The JOLINES fluxes are compared with SDSS spectroscopic measurements from the MPA--JHU DR8 catalogue (rescaled based on the SDSS aperture-corrected H$\alpha$ fluxes in \citealt{duartepuertas17}; blue circles), and DESI DR1 \citep{desi25} line fluxes from the stellar mass and emission line value-added catalogue (\citealt{zou24}; Zou et al. in prep.), corrected by the ratio between our \textit{i}-band \texttt{FLUX\_AUTO} fluxes and the $i$-band flux measured within the DESI fibre (green diamonds). In (b) we assume a $\sim 22\%$ contribution of the [\ion{N}{ii}]$\lambda 6583$ line to the H$\alpha$ + [\ion{N}{ii}] blend. Open symbols represent measurements with equivalent widths in the 5--20\,\AA\ range for the corresponding emission line, while filled symbols correspond to equivalent widths > 20\,\AA. In dashed grey style, we show the one-to-one lines. In all panels, the black contours represent the 1, 10, 50, 90, and 99\% percentiles of the two-dimensional probability distribution.
\end{figure*}

\subsection{Comparison with DESI synthetic photospectra}\label{sec_desi}
As a second test of our methodology, we used real galaxy spectra from the DESI EDR, to generate synthetic J-PAS observations. Specifically, we applied the J-PAS filter transmission curves to the DESI spectra. This procedure resulted in synthetic J-PAS photospectra, closely mimicking actual observations, but with the benefit of known spectroscopic line fluxes from DESI.

Subsequently, these synthetic photospectra were processed using the same approach described in Section\,\ref{obs}: we performed continuum fitting using CIGALE, subtracted the fitted continuum emission, and extracted the fluxes of the brightest emission lines listed in Table\,\ref{tab_lines}. The recovered emission-line fluxes were then directly compared with the spectroscopic fluxes provided by the FastSpecFit value-added catalogue (\citealt{moustakas23}; Moustakas et\,al. in prep.), associated with the DESI EDR.

Fig.\,\ref{fig_desi} illustrates the comparison between the synthetic photometric fluxes derived using our method and the original DESI spectroscopic measurements. The agreement between both datasets is excellent, demonstrating robust flux recovery for emission lines with equivalent widths exceeding EW\,$\gtrsim$20\,\AA, consistent with the performance observed in the simulated galaxies (Section\,\ref{sec_mock}). Notably, the ratio of recovered to spectroscopic fluxes in Fig.\,\ref{fig_desi} exhibits a significantly less pronounced bias at low equivalent widths than that seen in the mock catalogue (Fig.\,\ref{fig_mock}). We ascribe the more pronounced bias observed in the mock sample to the large difference in parameter values and continuum shapes between the mock and the SED fitting models. This mismatch limits the fitting model's ability to provide an accurate estimate of the continuum level, thus affecting flux measurements at low EW values. Conversely, the significantly lower deviation seen in the DESI comparison implies that the parameter space covered by the fitting models is well-suited to the characteristics of a real galaxy sample with varied spectral shapes. Therefore, the test based on synthetic photospectra from DESI confirms that our method for continuum subtraction and line flux extraction introduces no significant biases, validating the reliability and accuracy of the technique when applied to real observational data.

\subsection{Comparison with spectroscopic fluxes}\label{sec_spec}
After validating our method with the mock catalogue and the DESI synthetic photospectra, we searched for galaxies in the miniJPAS, J-NEP, and J-PAS EDR surveys with available spectroscopic flux measurements in public catalogues. Comparing photometric fluxes derived from narrow-band filters with spectroscopic measurements provides a direct assessment of the accuracy of our methodology. While variations in the apertures used for spectroscopic observations and narrow-band photometry can introduce differences in the line fluxes, these discrepancies are expected to be significant only for extended galaxies. For most compact sources, typical fibre or slit apertures wider than $\sim 2''$ should yield fluxes comparable to those measured by the $3''$ aperture photometry in miniJPAS, J-NEP, and J-PAS EDR.

Figures\,\ref{subfig_spec_Fo3} and \ref{subfig_spec_Fha} show a comparison between the emission-line fluxes derived from J-PAS photospectra, for galaxies with $i < 21\,\mathrm{mag}$, and spectroscopic measurements from three external datasets. First, we use the MPA--JHU DR8 value-added catalogue\footnote{\url{https://www.sdss4.org/dr17/spectro/galaxy_mpajhu}} of the Sloan Digital Sky Survey (SDSS) for galaxies with detected [\ion{O}{iii}]$\lambda4959,5007$ emission. These fluxes were rescaled using the ratio between the aperture-corrected H$\alpha$ flux from \citet{duartepuertas17} and the MPA--JHU DR8 H$\alpha$ measurement, thereby deriving an aperture correction that we then applied consistently to the rest of the SDSS emission lines. Second, we include measurements from the FastSpecFit value-added catalogue (\citealt{moustakas23}; Moustakas et al.\ in prep.) of the DESI Data Release~1 (DR1; \citealt{desi25}). Since FastSpecFit does not provide synthetic $i$-band fibre magnitudes, but does tabulate the $r$- and $z$-band fibre fluxes, we estimated the $i$-band flux using a first-degree power-law interpolation in wavelength. The DESI line fluxes were then corrected by the ratio between our \textit{i}-band \texttt{FLUX\_AUTO} fluxes and the interpolated $i$-band flux within the $1\farcs5$ DESI fibre. Finally, we compare with SDSS galaxies with aperture-corrected H$\alpha$ fluxes from \citet{duartepuertas17}, assuming no extinction correction and a $70\pm10\%$ contribution of H$\alpha$ to the H$\alpha$ + [\ion{N}{ii}] blend. The aperture corrections in \citet{duartepuertas17} are based on empirical relations derived from integral-field spectroscopic observations, providing a more reliable basis for comparison with the J-PAS photometric fluxes. Although this approach assumes that the spatial distribution of the line-emitting gas approximately follows that of the stellar continuum, this approximation has been shown to hold, at least to first order, for star-forming galaxies in large samples such as CALIFA \citep{iglesias-paramo13,iglesias-paramo16}, where the radial growth curves of H$\alpha$ and the $r$-band continuum show similar trends within the optical extent of the galaxy. This assumption remains valid when the fibre covers a significant fraction of the galaxy, typically above $\sim$20\,\% of its radius, as is the case for most SDSS galaxies at $z \gtrsim 0.04$ \citep{kewley05}. Under this approximation, the continuum-based aperture correction provides a practical means to estimate the total line flux from fibre spectra when no direct spatially resolved emission-line information is available. The lower panels in Fig.\,\ref{fig_spec} show the J-PAS-to-spectroscopic line flux ratios as a function of the equivalent width of [\ion{O}{iii}]$\lambda 5007$ (Fig.\,\ref{subfig_spec_ewo3}) or H$\alpha$ (Fig.\,\ref{subfig_spec_ewha}) for a common sample of 1302 and 4060 galaxies, respectively. In all cases, open symbols correspond to measurements with EW([\ion{O}{iii}]$\lambda 5007$) (Figs.\,\ref{subfig_spec_Fo3} and \ref{subfig_spec_ewo3}) or EW(H$\alpha$) (Figs.\,\ref{subfig_spec_Fha} and \ref{subfig_spec_ewha}) in the 5--20\,\AA\ range, while filled symbols correspond to galaxies with EW values above 20\,\AA.

Despite the relatively large uncertainties and scatter in Figs\,\ref{subfig_spec_Fo3} and \ref{subfig_spec_ewo3}, the observed trend for the majority of galaxies with bright [\ion{O}{iii}]$\lambda 4959,5007$ detections suggests a good agreement between the J-PAS spectrophotometric fluxes and the spectroscopic values, especially at large EW([\ion{O}{iii}]$\lambda 5007$) values. This is confirmed by the larger sample of common sources with H$\alpha$ + [\ion{N}{ii}]$\lambda 6548,6583$ measurements in Figs\,\ref{subfig_spec_Fha} and \ref{subfig_spec_ewha}. Overall, galaxies with spectroscopic fluxes above $\gtrsim 2 \times 10^{-15}\, \rm{erg\,s^{-1}\,cm^{-2}}$ show a good agreement with J-PAS measurements, with a significant decrease in the scatter for sources with EW(H$\alpha$)$ > 20$\,\AA\ (filled symbols), as predicted by results obtained for the simulated photospectra (Section\,\ref{sec_mock}). A small bias towards larger J-PAS fluxes is noticeable for galaxies with EWs in the 5--20\,\AA\ range, which is partially attributable to uncorrected differences in the flux extraction apertures.

In Figure\,\ref{fig_spec_faint}, we show the ability of our method to recover the fluxes of other relevant lines in the optical spectrum, namely the [\ion{O}{ii}]$\lambda 3726,3729$ and [\ion{S}{ii}]$\lambda 6716,6731$ doublets, and the [\ion{N}{ii}]$\lambda 6583$ line. The latter was derived from the filters containing the H$\alpha$ + [\ion{N}{ii}]$\lambda 6548$ and H$\alpha$ + [\ion{N}{ii}]$\lambda 6583$ blends, by assuming a $\sim$22\,\% contribution from the [\ion{N}{ii}]$\lambda 6583$ line to the total H$\alpha$ + [\ion{N}{ii}]$\lambda 6548,6583$ flux, and adopting the theoretical [\ion{N}{ii}]$\lambda 6583$/[\ion{N}{ii}]$\lambda 6548$ ratio of $2.94$ \citep[for $T_\text{e}=10^4\,\rm{K}$ and $n_\text{e}=100\,\rm{cm^{-3}}$;][]{luridiana15}. In all three cases, the emission-line fluxes are successfully recovered, with a small bias towards slightly higher fluxes in J-PAS at lower flux or equivalent width values. The median and dispersion of the ratios between the J-PAS and spectroscopic flux measurements are $0.8 \pm 0.2$ for [\ion{O}{ii}], $1.5 \pm 0.7$ for [\ion{N}{ii}], and $1.0 \pm 0.3$ for [\ion{S}{ii}], indicating a generally good agreement within the expected uncertainties. The successful recovery of the [\ion{N}{ii}]$\lambda6583$ flux in this figure demonstrates the capability of the J-PAS narrow-band system to separate blended features and recover individual emission-line fluxes. This is achieved by combining the information from adjacent filters and relying on only minimal theoretical assumptions, highlighting the potential of the method to retrieve, for example, H$\alpha$ fluxes decontaminated from [\ion{N}{ii}] or individual [\ion{O}{iii}] and H$\beta$ fluxes. This implies that robust galaxy selections based on these emission lines can be performed, enabling the construction of large samples of galaxies with, for example, high N/O ratios using the [\ion{N}{ii}]$\lambda 6583$/[\ion{O}{ii}]$\lambda 3727$ ratio, strong ionising radiation fields traced by the [\ion{O}{iii}]$\lambda 5007$/[\ion{O}{ii}]$\lambda3727$ ratio, or active nuclei identified through diagnostic ratios such as [\ion{O}{iii}]$\lambda 5007$/H$\beta$ and [\ion{S}{ii}]$\lambda 6716,6731$/H$\alpha$.

Overall, the comparison with line spectroscopic fluxes suggests that the continuum-subtraction method based on SED fitting described in Section\,\ref{obs} is capable of reliably recovering emission-line fluxes from galaxy photospectra in miniJPAS, J-NEP, and J-PAS EDR. These results further validate the robustness of the photometric approach for detecting and measuring emission lines in compact galaxies with sufficient flux strength. See Section\,\ref{discuss} for a detailed discussion on the overall properties of emission-line galaxies included in the J-PAS optical line intensities for nebular emission galaxies (JOLINES) catalogue and the implications of these results for future surveys.

\begin{figure*}[t]
  \centering
  \subfigure[]{\includegraphics[width = 0.5\textwidth]{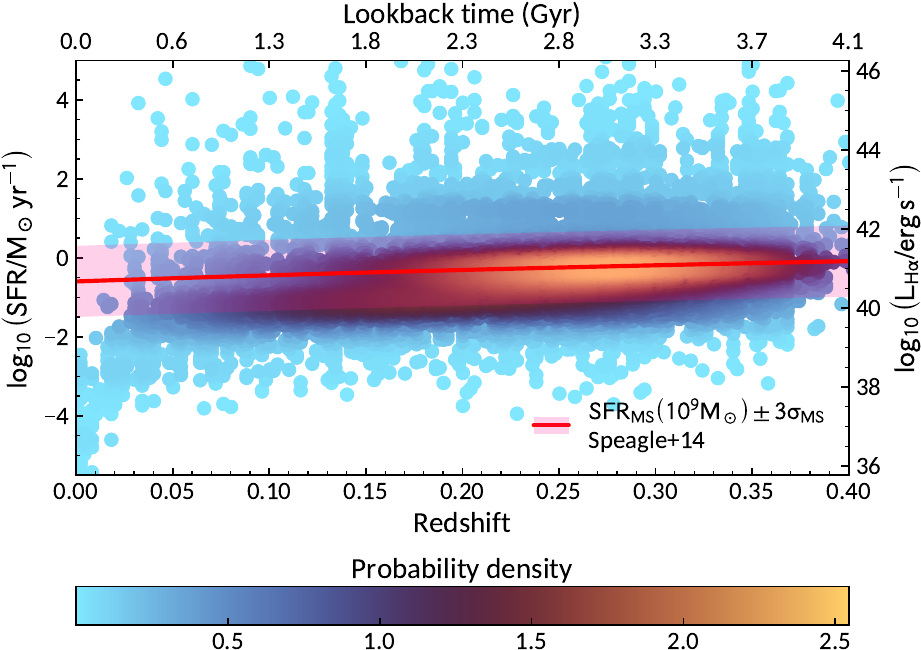}\label{subfig_SFR}}~
  \subfigure[]{\includegraphics[width = 0.5\textwidth]{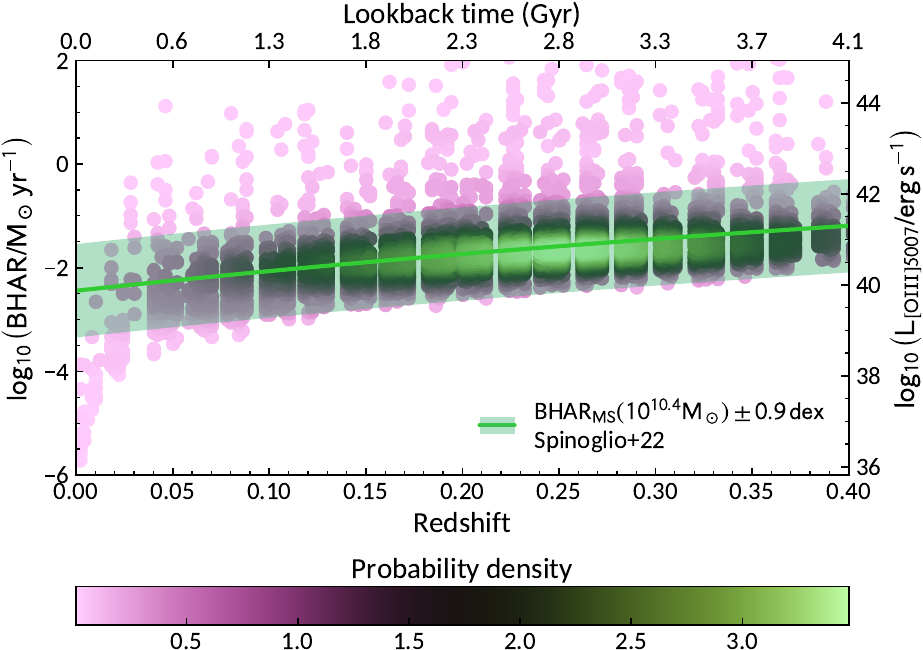}\label{subfig_BHAR}}
  \caption{Distribution of SFR as a function of redshift (a) for galaxies detected in miniJPAS, J-NEP, and J-PAS EDR, determined from their extinction-corrected H$\alpha$ emission \citep{kennicutt12}. The evolution of the main sequence for a galaxy with $M_* = 10^9\, \rm{M_\odot}$ and its associated dispersion are shown for comparison \citep[solid line and shaded area in red colour;][]{speagle14}. Evolution of BHAR as a function of redshift (b), derived from the extinction-corrected [\ion{O}{iii}]$\lambda5007$ luminosity \citep{spinoglio24}. The SFR--BHAR relation for an AGN galaxy with $M_* = 10^{10.4}\, \rm{M_\odot}$, obtained for AGN in the nearby Universe, is also shown \citep[solid line and shaded area in green colour;][]{spinoglio22}.}\label{fig_Llines}
\end{figure*}

\section{Emission-line flux catalogue}\label{vac}

In this section, we provide a description of the JOLINES value-added catalogue, which contains emission-line fluxes for all galaxies in the miniJPAS, J-NEP, and J-PAS EDR surveys in its first release. The catalogue serves as a foundational resource for analysing emission-line galaxies within these surveys and is designed to be expanded in future releases. As the J-PAS survey progresses, JOLINES will be updated to include emission-line fluxes derived from the photospectra of galaxies observed in forthcoming data releases, ensuring the catalogue remains a comprehensive and up-to-date tool for galaxy evolution studies.

The JOLINES catalogue is structured to provide key information for each galaxy, including the \textsc{tile-id} and \textsc{number} identifiers assigned by the miniJPAS, J-NEP, or J-PAS EDR catalogues, the galaxy coordinates in right ascension and declination, as well as the spectrophotometric redshifts. Additionally, it includes the emission-line fluxes or line blends available for each target, and their associated uncertainties. Certain line blends, such as H$\beta$ + [\ion{O}{iii}]$\lambda 4959$, are only present at specific redshifts, depending on the filter configuration and the rest-frame wavelengths of the lines. When possible, the catalogue provides individual line fluxes, which are disentangled using theoretical line ratios and fluxes from adjacent filters, as described in Section\,\ref{obs}. This ensures that the information is as complete and accurate as possible for a wide range of galaxies and redshifts.

For cases where individual line fluxes are measured for lines close in wavelength, that may be blended for other targets at different redshift such as H$\beta$ + [\ion{O}{iii}]$\lambda 4959$, the catalogue also includes the total line blend flux for completeness. Flux measurements from filters contaminated by adjacent emission lines in their transmission wings have been corrected. However, these measurements are flagged to indicate possible large uncertainties. These are primarily caused by wavelength shifts in the transmission profiles (of the order of 10--15\,\AA) due to spatial variations across the field of view, which can affect the net transmission for emission lines in the filter side.
 
The current JOLINES catalogue, covering miniJPAS, J-NEP, and J-PAS EDR, comprises approximately 13\,900 sources with significant emission ($> 3\sigma$) in the H$\alpha$+[\ion{N}{ii}]$\lambda 6548,6583$ blend. Additionally, the catalogue includes 6\,100 detections in [\ion{O}{ii}]$\lambda 3727$, 1\,700 in [\ion{Ne}{iii}]$\lambda 3869$, 2\,900 in H$\beta$, 7\,200 sources in [\ion{O}{iii}]$\lambda 5007$, 500 in [\ion{O}{i}]$\lambda 6300$, 6\,100 in H$\alpha$, and 4\,000 in [\ion{S}{ii}]$\lambda 6724$. For a comprehensive list and description of all data columns included in the JOLINES catalogue, refer to Table\,\ref{tab_lines} in Appendix\,\ref{app_jlines}. The JOLINES catalogue can be found in the online version of this publication and is also distributed as a value-added product alongside the official data releases of the miniJPAS\footnote{\url{https://www.j-pas.org/datareleases/minijpas_public_data_release_pdr201912}}, J-NEP\footnote{\url{https://archive.cefca.es/catalogues/jnep-pdr202107}}, and the J-PAS EDR\footnote{\url{https://archive.cefca.es/catalogues/jpas-edr}} surveys. It can be accessed through their respective data portals, providing an easily available resource for the astronomical community.

\section{Discussion}\label{discuss}

As shown by the analysis in Section\,\ref{results}, the JOLINES catalogue comprises emission-line galaxies with at least one bright emission line exhibiting an equivalent width of $\gtrsim 10$--$20$\,\AA\ within the redshift range $0 < z < 0.4$. This threshold predominantly selects star-forming galaxies, AGN, and composite sources, defining a sample that is particularly well suited for studying the evolution of these phenomena in the local Universe \citep[e.g.][]{gonzalez-delgado22,rodriguez-martin22,lopez23}. In this section, our aim is to provide a general overview of the expected properties for the sample of galaxies included in the JOLINES catalogue.

The most prominent lines detected for star-forming galaxies and AGN are typically H$\alpha$ and [\ion{O}{iii}]$\lambda 5007$, respectively, which show median luminosities and associated dispersions of $\log(L_{\rm H\alpha} / \rm{erg\,s^{-1}}) \sim 40.8 \pm 0.9$ and $\log(L_\text{[\ion{O}{iii}]5007} / \rm{erg\,s^{-1}}) \sim 40.8 \pm 0.6$. These have been corrected for dust extinction using the Balmer decrement (H$\alpha$/H$\beta$) measured in JOLINES, assuming an intrinsic ratio of 2.87, consistent with Case B recombination in \citet{osterbrock06}. To distinguish star-forming galaxies from those with a significant AGN contribution, we classify as AGN those sources with EW([\ion{O}{iii}]$\lambda5007$)$\,\geq 5\,\rm{\mathring{A}}$ and satisfying at least one of the following conditions: [\ion{O}{iii}]$\lambda5007$/H$\beta \geq 5$, [\ion{S}{ii}]$\lambda6725$/H$\alpha \geq 0.55$, [\ion{O}{i}]$\lambda6300$/H$\alpha \geq 0.08$, or [\ion{O}{iii}]$\lambda5007$/[\ion{O}{ii}]$\lambda3727 \geq 2$ \citep[e.g.][]{kewley06,stasinska25}. We note that AGN continuum emission has not been included in the SED fitting, however, only powerful Type 1 AGN are expected to noticeably affect the continuum in the wavelength and redshift range probed here. In such cases the continuum would appear as a bluer power-law, which the model would compensate by adding a larger fraction of young, blue stars, but the overall continuum level would remain comparable, and the resulting emission-line flux measurements would not be significantly affected (see Fig.\,\ref{fig_desi_agn} and discussion in Appendix\,\ref{app_agn}).

An EW threshold of $\gtrsim$10--20\,\AA\ in H$\alpha$ strongly favours galaxies undergoing recent star formation episodes, typically within the last $\lesssim 10\, \rm{Myr}$, with SFR in the $0.1$--$2\, \rm{M_\odot\,yr^{-1}}$ range, as estimated from the median extinction-corrected H$\alpha$ luminosity and its associated dispersion in our sample \citep{kennicutt12}. These values are characteristic of main-sequence galaxies with stellar masses close to the knee of the stellar mass function (\citealt{noeske07,elbaz07,elbaz11,duartepuertas17}). In this regard, Fig.\,\ref{subfig_SFR} shows the distribution of SFR in our sample as a function of redshift, excluding the AGN population. The general trend aligns with the star-forming main sequence evolution for galaxies with a stellar mass around $M_* = 10^9\, \rm{M_\odot}$ \citep[solid line and shaded area in red;][]{speagle14}, suggesting that JOLINES predominantly captures galaxies actively forming stars. This is in agreement with the results obtained by \citet{martinez-solaeche22}, who employed machine learning techniques to analyse the miniJPAS sample \citep[see also results by][based on spectroscopic data]{favole24}. At higher redshifts, approaching $z \sim 0.4$, the general increase in SFR and correspondingly higher EW(H$\alpha$) values \citep[e.g.][]{fumagalli12} facilitate the detection of slightly less massive galaxies in JOLINES. Additionally, the catalogue includes a small fraction of dwarf galaxies with elevated sSFR ($\gtrsim 3\, \rm{Gyr^{-1}}$) and exceptionally high EW([\ion{O}{iii}]$\lambda 5007$) or EW(H$\alpha$) values of $\gtrsim 300$\,\AA\ (\citealt{iglesias-paramo22,lumbreras-calle22}; Giménez-Alcázar in prep.), which are often associated with intense starburst episodes in low-metallicity galaxies.

On the other hand, the [\ion{O}{iii}]$\lambda 5007$ line allows us to detect a significant fraction of the Seyfert galaxy population at low redshift, which exhibits a characteristic luminosity of $\log(L^*_\text{[\ion{O}{iii}]5007} / \rm{erg\,s^{-1}}) \sim 41.4$ at the knee of the [\ion{O}{iii}]$\lambda 5007$ luminosity function \citep{buongiorno10}. Figure\,\ref{subfig_BHAR} shows the black hole accretion rate (BHAR) as a function of redshift for an active galaxy with $M_* = 10^{10.4}\, \rm{M_\odot}$, derived from the BHAR--SFR correlation found in nearby AGN \citep{spinoglio22}, using the extinction-corrected [\ion{O}{iii}]$\lambda5007$ luminosity as a proxy for the BHAR \citep{spinoglio24}:
\begin{equation}
    \frac{\mathrm{BHAR}}{\mathrm{M_\odot\,yr^{-1}}} = \frac{L_\mathrm{bol}}{\eta\,c^2} \approx 10^{-36.5 \pm 0.2} \times \left(\frac{L_\text{[\ion{O}{iii}]5007}}{\mathrm{erg\,s^{-1}}}\right)^{0.88 \pm 0.05}\\
\end{equation}
assuming a radiative efficiency of $\eta = 0.1$. The observed distribution of $\log(L_\text{[\ion{O}{iii}]5007})$ with redshift in Fig.\,\ref{subfig_BHAR} indicates that JOLINES includes a significant fraction of moderately luminous AGN, consistent with known relationships between BHAR and SFR in local active galaxies.

Finally, we highlight that other galaxy populations, such as low-luminosity AGN, numerous in the nearby Universe \citep{ho08,burke25}, are expected to be detected through their characteristic strong low-ionisation emission lines \citep{heckman80}. A rough comparison between the median extinction-corrected luminosities for transitions from low-ionisation species in JOLINES (e.g. $\log(L_\text{[\ion{O}{ii}]3727} / \rm{erg\,s^{-1}}) \sim 40.9 \pm 0.5$) and the characteristic line luminosities at the knee of the luminosity functions for these nuclei ($\log(L^*_\text{[\ion{O}{ii}]3727} / \rm{erg\,s^{-1}}) \sim 40.5$;  \citealt{favole24}), suggests that JOLINES is also expected to include a significant fraction of low-luminosity AGN.

These findings underscore the capability of JOLINES to capture a diverse population of galaxies, spanning a wide range of star formation and AGN activity, thereby providing a valuable resource for studying galaxy evolution at low redshift.

% log(Lbol/ 1.e+41 erg/s) = (log(Lo3 / 1.e+41 erg/s) + 3.76) / 1.14

%%%%%%%%%%%%%%%%%%%%%%%%%%%%%%%%%%%%%%%%%%%%%%%%%%%%%%%%%%%%%%
\section{Summary}\label{summary}

%Reliable fluxes can be obtained for emission lines with equivalent width values above EW > 20 \AA \, providing Halpha+[NII] fluxes for the majority of galaxies in the main sequence at z < 0.4. Likewise, [OIII]5007\AA \ fluxes can be measured for a large fraction of Seyfert galaxies at low-z. The first catalogue release includes emission line fluxes for galaxies in miniJPAS and J-NEP. This value-added catalogue will be publicly available as part of the J-PAS survey data releases.

In this study, we introduced the JOLINES value-added catalogue, a dataset containing emission-line fluxes for galaxies in the miniJPAS and J-NEP precursor surveys, and in the J-PAS EDR. The catalogue is constructed using a robust SED-fitting approach with CIGALE, which allows accurate determination of the continuum emission and precise measurement of line fluxes from narrow-band photometry. This methodology is crucial for obtaining reliable emission-line fluxes, as direct continuum subtraction can introduce significant uncertainties, particularly for faint lines.

The accuracy of our method was tested using simulated galaxy spectra with added noise, allowing us to assess how well individual and blended emission-line fluxes can be recovered. We find that our approach successfully retrieves line fluxes with minimal bias, particularly for lines with equivalent widths EW\,$\gtrsim$20\,\AA. Additionally, we performed a comparison with spectroscopic measurements from SDSS and DESI, confirming that emission-line fluxes derived from J-PAS photospectra are consistent with spectroscopic values, with a typical dispersion of $\sim$0.3\,dex for bright lines. The agreement improves for stronger emission lines, reinforcing the reliability of the method for star-forming and AGN-host galaxies.

The JOLINES catalogue includes galaxy identifiers, sky coordinates, spectro-photometric redshifts, and emission-line fluxes or blends for each target. Line blends, such as H$\beta$ + [\ion{O}{iii}]$\lambda4959$, are included when present, and individual line fluxes are provided where possible using theoretical line ratios and measurements from adjacent filters. Flux measurements affected by contamination from nearby lines in the filter transmission wings are corrected but remain flagged due to residual uncertainties. The first JOLINES release provides statistically robust samples for the brightest optical emission lines, including approximately 13\,900 galaxies with significant detections in the H$\alpha$+[\ion{N}{ii}] complex, 7\,200 in [\ion{O}{iii}]$\lambda5007$, 6\,100 in [\ion{O}{ii}]$\lambda3727$, 2\,900 in H$\beta$, and 4\,000 in [\ion{S}{ii}]$\lambda6716,6731$, among others. These large samples enable population-wide studies across the full sample of galaxies in J-PAS surveys.

This catalogue is publicly available as part of the miniJPAS, J-NEP, and J-PAS EDR data releases and will be continuously updated with future J-PAS data. By providing a robust dataset for emission-line galaxies, JOLINES will enable statistical studies of star formation, AGN activity, and the interstellar medium across a wide range of galaxy populations.

%%%%%%%%%%%%%%%%%%%%%%%%%%%%%%%%%%%%%%%%%%%%%%%%%%%%%%%%%%%%%%
\begin{acknowledgements}
%The authors would like to thank the anonymous referee for his/her useful comments and suggestions.
Based on observations made with the JST250 telescope and JPCam at the Observatorio Astrof\'{\i}sico de Javalambre (OAJ), in Teruel, owned, managed, and operated by the Centro de Estudios de F\'{\i}sica del Cosmos de Arag\'on (CEFCA). This paper has gone through internal review by the J-PAS collaboration. JAFO acknowledges financial support by the Spanish Ministry of Science and Innovation (MCIN/AEI/10.13039/501100011033), by ``ERDF A way of making Europe'' and by ``European Union NextGenerationEU/PRTR'' through the grants PID2021-124918NB-C44 and CNS2023-145339; MCIN and the European Union -- NextGenerationEU through the Recovery and Resilience Facility project ICTS-MRR-2021-03-CEFCA. RMGD acknowledges financial support from the Severo Ochoa grant CEX2021-001131-S funded by MICIU/AEI/ 10.13039/501100011033, and from the project PID2022-141755NB-I00. RA, AHC and JZC, acknowledge support of grant PID2023-147386NB-I00, funded by MICIU/AEI/10.13039/501100011033 and by ERDF/EU. IB has received funding from the European Union's Horizon 2020 research and innovation programme under the Marie Skłodowska--Curie Grant Agreement No.\ 101059532. 
We acknowledge the OAJ Data Processing and Archiving Department (DPAD) for reducing and calibrating the OAJ data used in this work. Funding for the J-PAS Project has been provided by the Governments of Spain and Arag\'on through the Fondo de Inversiones de Teruel; the Aragonese Government through the Research Groups E96, E103, E16\_17R, E16\_20R, and E16\_23R; the Spanish Ministry of Science and Innovation (MCIN/AEI/10.13039/501100011033 and ``ERDF A way of making Europe'') with grants PID2021-124918NB-C41, PID2021-124918NB-C42, PID2021-124918NA-C43, and PID2021-124918NB-C44; the Spanish Ministry of Science, Innovation and Universities (MCIU/AEI/FEDER, UE) with grants PGC2018-097585-B-C21 and PGC2018-097585-B-C22; the Spanish Ministry of Economy and Competitiveness (MINECO) under AYA2015-66211-C2-1-P, AYA2015-66211-C2-2, and AYA2012-30789; and European FEDER funding (FCDD10-4E-867, FCDD13-4E-2685). The Brazilian agencies FINEP, FAPESP, FAPERJ and the National Observatory of Brazil have also contributed to this project. Additional funding was provided by the Tartu Observatory and by the J-PAS Chinese Astronomical Consortium. This research has made use of the SVO Filter Profile Service ``Carlos Rodrigo'', funded by MCIN/AEI/10.13039/501100011033/ through grant PID2020-112949GB-I00. This work made use of \textsc{Astropy}\footnote{\url{http://www.astropy.org}}: a community-developed core Python package and an ecosystem of tools and resources for astronomy \citep{astropy22}.
\end{acknowledgements}

%%%%%%%%%%%%%%%%%%%%%%%%%%%%%%%%%%%%%%%%%%%%%%%%%%%%%%%%%%%%%%
\bibliographystyle{aa}
\bibliography{jolines}
%%%%%%%%%%%%%%%%%%%%%%%%%%%%%%%%%%%%%%%%%%%%%%%%%%%%%%%%%%%%%%
\begin{appendix}
\onecolumn
\begin{table}[t!]
\section{Mock catalogue parameters}\label{app_mock}
\caption{Parameters and their respective sampling ranges for the various modules used to generate the mock catalogue with CIGALE, used in Section\,\ref{results} to test the reliability of the method developed to obtain emission-line fluxes from the narrow-band spectrophotometic data.}
\centering
\setlength{\tabcolsep}{1.mm}
\begin{tabular}{cc}
\bf Parameters &  \bf Sampling range\\
\hline\\[-1.5ex]
\multicolumn{2}{c}{\it Star formation history: delayed model} \\[0.5ex]
Age of the main population &  7, 8, 9 Gyr  \\
e-folding time & 1, 2 Gyr \\ 
Age of the young burst &  8, 20, 80, 150, 300, 500 Myr  \\
e-folding time young burst & 2, 5 Myr \\ 
Burst stellar mass fraction & 0.01, 0.05, 0.1 \\
\hline\\[-1.5ex]
\multicolumn{2}{c}{\it Simple Stellar populations: \citet{bruzual03}} \\[0.5ex]
Initial Mass Function & \citet{salpeter55} \\
Metallicity & 0.004, 0.008, 0.02, 0.05 \\
\hline\\[-1.5ex]
\multicolumn{2}{c}{\it Nebular emission} \\[0.5ex]
Ionisation parameter (log\,U) & -3.0, -3.3, -3.5, -3.7, -4.0 \\
Gas metallicity ($Z_\text{gas}$) & 0.001, 0.004, 0.014, 0.022, 0.041 \\
Electron density ($\rm n_e$)     & 100 cm$^{-3}$ \\
\hline\\[-1.5ex]
\multicolumn{2}{c}{\it Dust extinction} \\[0.5ex]
Dust attenuation law & modified \citet{calzetti00}\\
E(B-V)$_\text{lines}$ colour excess of the nebular lines light & 0.0, 0.1, 0.2, 0.3, 0.4, 0.5, 0.6, 0.7, 0.8 \\
Continuum attenuation E(B-V)$_\text{stellar}$ fraction relative to E(B-V)$_\text{lines}$ & 0.44 \\
Ratio of total to selective extinction $\rm A_V / E(B-V)$ & 3.1 \\
%Amplitude of the UV bump & 0 \\
%Slope delta of the power law modifying the attenuation curve & 0 \\
Extinction law & \citet{cardelli89} \\
\hline\\[-1.5ex]
Redshift values & 0.01--0.40 (step 0.01) \\
\hline\\[-1.5ex]
Number of models per redshift & 194\,400 \\[0.5ex]
\hline
\label{tab_mockpars}
\end{tabular}
\end{table}

\begin{table*}
\section{Catalogue description}\label{app_jlines}

\caption[]{Catalogue keywords for the JOLINES catalogue. All emission line fluxes are provided in units of $10^{-15}\, \rm{erg\,s^{-1}\,cm^{-2}}$.}
\small
\centering
\begin{tabular}{ll}\label{tab:statistics}
  \textbf{Keyword}       &  \textbf{Description} \\
  \hline\\[-0.3cm]
  \texttt{TILE\_ID}  & \text{Tile ID of the reference $r$ band (miniJPAS and J-NEP) or $i$ band (J-PAS EDR)} \\[0.1cm]
  \texttt{NUMBER}  & \text{Number id assigned by \texttt{SExtractor} to the object in J-PAS.} \\[0.1cm]
  \texttt{R.A.}  & \text{Right ascension (J2000)}\\[0.1cm]
  \texttt{Dec.}  & \text{Declination (J2000)}\\[0.1cm]
  \texttt{PhotoZ}  & \text{Spectrophotometric redshift}\\[0.1cm]
  \texttt{o2\_3727}  & \text{[\ion{O}{ii}]$\lambda3727$ emission-line doublet flux}\\[0.1cm]
  \texttt{err\_3727}  & \text{[\ion{O}{ii}]$\lambda3727$ emission-line doublet flux uncertainty} \\[0.1cm]
  \texttt{ew\_3727}  & \text{[\ion{O}{ii}]$\lambda3727$ emission-line doublet equivalent width} \\[0.1cm]
  \texttt{ne3\_3869}  & \text{[\ion{Ne}{iii}]$\lambda3869$ emission-line flux}\\[0.1cm]
  \texttt{err\_3869}  & \text{[\ion{Ne}{iii}]$\lambda3869$ emission-line flux uncertainty} \\[0.1cm]
  \texttt{ew\_3869}  & \text{[\ion{Ne}{iii}]$\lambda3869$ emission-line flux equivalent width}\\[0.1cm]
  \texttt{hg+o3\_4352}  & \text{H$\gamma$ $\lambda4340$ + [\ion{O}{iii}]$\lambda4363$ emission-line flux of the blend}\\[0.1cm]
  \texttt{err\_4352}  & \text{H$\gamma$ $\lambda4340$ + [\ion{O}{iii}]$\lambda4363$ emission-line flux uncertainty of the blend} \\[0.1cm]
  \texttt{ew\_4352}  & \text{H$\gamma$ $\lambda4340$ + [\ion{O}{iii}]$\lambda4363$ emission-line equivalent width of the blend} \\[0.1cm]
  \texttt{hb\_4861}  & \text{H$\beta$ $\lambda4861$ emission-line flux} \\[0.1cm]
  \texttt{err\_4861}  & \text{H$\beta$ $\lambda4861$ emission-line flux uncertainty} \\[0.1cm]
  \texttt{ew\_4861}  & \text{H$\beta$ $\lambda4861$ emission-line equivalent width} \\[0.1cm]
  \texttt{o3\_4959}  & \text{[\ion{O}{iii}]$\lambda4959$ emission-line flux}\\[0.1cm]
  \texttt{err\_4959} & \text{[\ion{O}{iii}]$\lambda4959$ emission-line flux uncertainty}\\[0.1cm]
  \texttt{o3\_5007}  & \text{[\ion{O}{iii}]$\lambda5007$ emission-line flux}\\[0.1cm]
  \texttt{err\_5007} & \text{[\ion{O}{iii}]$\lambda5007$ emission-line flux uncertainty}\\[0.1cm]
  \texttt{ew\_5007}  & \text{[\ion{O}{iii}]$\lambda5007$ emission-line equivalent width}\\[0.1cm]
  \texttt{hb+o3\_4910}  & \text{H$\beta$ $\lambda4861$ + [\ion{O}{iii}]$\lambda4959$ emission-line flux of the blend} \\[0.1cm]
  \texttt{err\_4910}  & \text{H$\beta$ $\lambda4861$ + [\ion{O}{iii}]$\lambda4959$ emission-line doublet flux uncertainty} \\[0.1cm]  
  \texttt{o3\_4983}  & \text{[\ion{O}{iii}]$\lambda4959,5007$ emission-line doublet flux} \\[0.1cm]
  \texttt{err\_4983}  & \text{[\ion{O}{iii}]$\lambda4959,5007$ emission-line doublet flux uncertainty} \\[0.1cm]
  \texttt{o1\_6300}  & \text{[\ion{O}{i}]$\lambda6300$ emission-line flux}\\[0.1cm]
  \texttt{err\_6300} & \text{[\ion{O}{i}]$\lambda6300$ emission-line flux uncertainty}\\[0.1cm]
  \texttt{ew\_6300}  & \text{[\ion{O}{i}]$\lambda6300$ emission-line equivalent width}\\[0.1cm]
  \texttt{ha\_6563} & \text{H$\alpha$ $\lambda6563$ emission-line flux of the line} \\[0.1cm]
  \texttt{err\_6563} & \text{H$\alpha$ $\lambda6563$ emission-line flux uncertainty of the line} \\[0.1cm]
  \texttt{ew\_6563} & \text{H$\alpha$ $\lambda6563$ emission-line equivalent width} \\[0.1cm]
  \texttt{n2\_6548} & \text{[\ion{N}{ii}]$\lambda6548$ emission-line flux of the line} \\[0.1cm]
  \texttt{err\_6548} & \text{[\ion{N}{ii}]$\lambda6548$ emission-line flux uncertainty of the line} \\[0.1cm]
  \texttt{n2\_6583} & \text{[\ion{N}{ii}]$\lambda6583$ emission-line flux of the line} \\[0.1cm]
  \texttt{err\_6583} & \text{[\ion{N}{ii}]$\lambda6583$ emission-line flux uncertainty of the line} \\[0.1cm]
  \texttt{ew\_6583} & \text{[\ion{N}{ii}]$\lambda6583$ emission-line equivalent width} \\[0.1cm]
  \texttt{n2+ha\_6555}  & \text{[\ion{N}{ii}]$\lambda6548$ + H$\alpha$ $\lambda6563$ emission-line flux of the blend} \\[0.1cm]
  \texttt{err\_6555}  & \text{[\ion{N}{ii}]$\lambda6548$ + H$\alpha$ $\lambda6563$ emission-line flux uncertainty of the blend} \\[0.1cm]
  \texttt{ha+n2\_6565} & \text{H$\alpha$ $\lambda6563$ + [\ion{N}{ii}]$\lambda6548,6583$ emission-line flux of the blend} \\[0.1cm]
  \texttt{err\_6565} & \text{H$\alpha$ $\lambda6563$ + [\ion{N}{ii}]$\lambda6548,6583$ emission-line flux uncertainty of the blend} \\[0.1cm]
  \texttt{ew\_6565} & \text{H$\alpha$ $\lambda6563$ + [\ion{N}{ii}]$\lambda6548,6583$ emission-line equivalent width of the blend} \\[0.1cm]
  \texttt{ha+n2\_6573}  & \text{H$\alpha$ $\lambda6563$ + [\ion{N}{ii}]$\lambda6583$ emission-line flux of the blend} \\[0.1cm]
  \texttt{err\_6573} & \text{H$\alpha$ $\lambda6563$ + [\ion{N}{ii}]$\lambda6583$ emission-line flux uncertainty of the blend} \\[0.1cm]
  \texttt{s2\_6724}  & \text{[\ion{S}{ii}]$\lambda6716,6731$ emission-line doublet flux} \\
  \texttt{err\_6724} & \text{[\ion{S}{ii}]$\lambda6716,6731$ emission-line doublet flux uncertainty} \\
  \texttt{ew\_6724}  & \text{[\ion{S}{ii}]$\lambda6716,6731$ emission-line doublet equivalent width} \\
  \texttt{s3\_9069}  & \text{[\ion{S}{iii}]$\lambda9069$ emission-line flux} \\
  \texttt{err\_9069} & \text{[\ion{S}{iii}]$\lambda9069$ emission-line flux uncertainty} \\
  \texttt{ew\_9069}  & \text{[\ion{S}{iii}]$\lambda9069$ emission-line equivalent width} \\
  \texttt{s3\_9531}  & \text{[\ion{S}{iii}]$\lambda9531$ emission-line flux} \\
  \texttt{err\_9531} & \text{[\ion{S}{iii}]$\lambda9531$ emission-line flux uncertainty} \\
  ... & ... \\[0.1cm]
  \hline
\end{tabular}
\end{table*}

\clearpage

\section{AGN population}\label{app_agn}

In this section, we analyse synthetic J-PAS photospectra generated for the AGN population in the DESI Early Data Release (EDR) \citep{desi24}. AGN were selected using the spectroscopic fluxes provided by the FastSpecFit value-added catalogue (\citealt{moustakas23}; Moustakas et al. in prep.), associated with the DESI EDR, and the classical AGN demarcation in the BPT diagram \citep{baldwin81}, based on the [\ion{O}{iii}]$\lambda 5007$/H$\beta$ and [\ion{N}{ii}]$\lambda 6583$/H$\alpha$ line ratios \citep{kewley01}.

\begin{figure*}[h!!!]
  \centering
  \subfigure[]{\includegraphics[width = 0.49\textwidth]{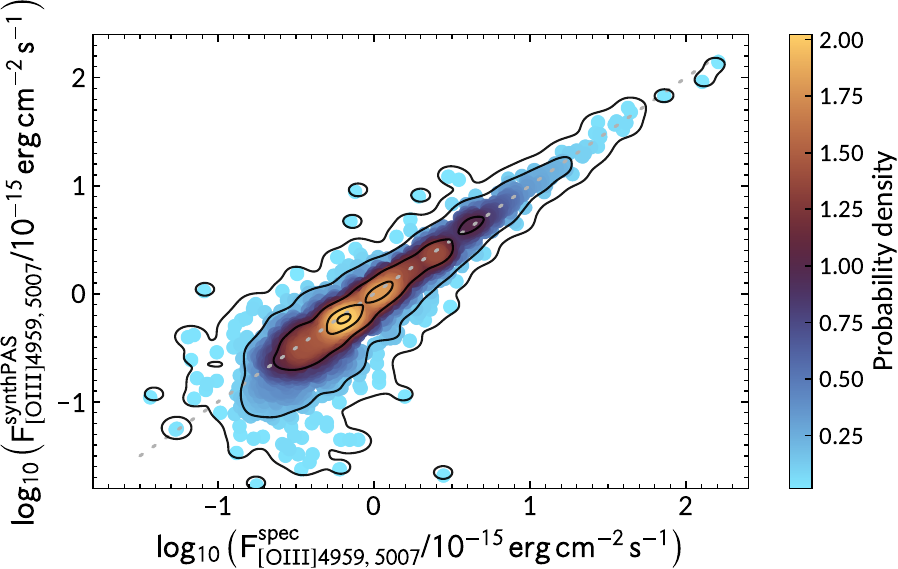}\label{subfig_desi_agn_Fo3}}~
  \subfigure[]{\includegraphics[width = 0.49\textwidth]{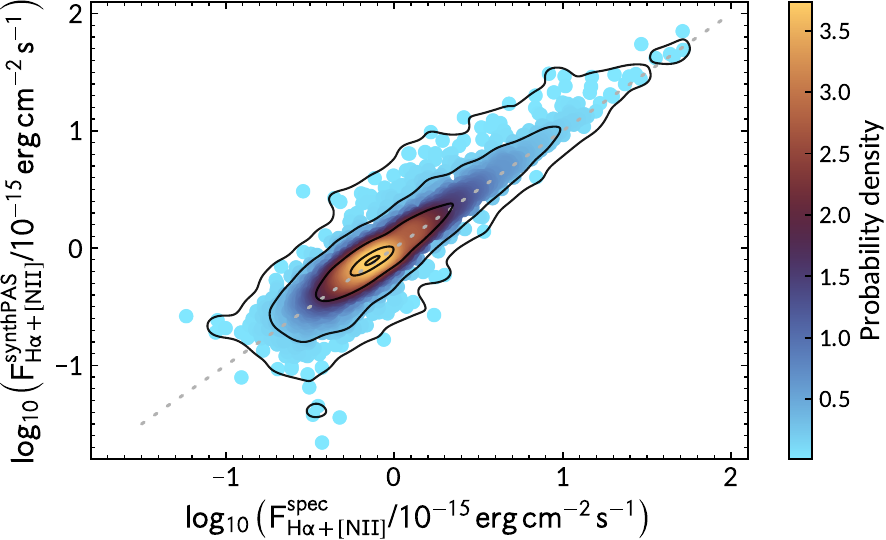}\label{subfig_desi_agn_Fha}}
  \subfigure[]{\includegraphics[width = 0.49\textwidth]{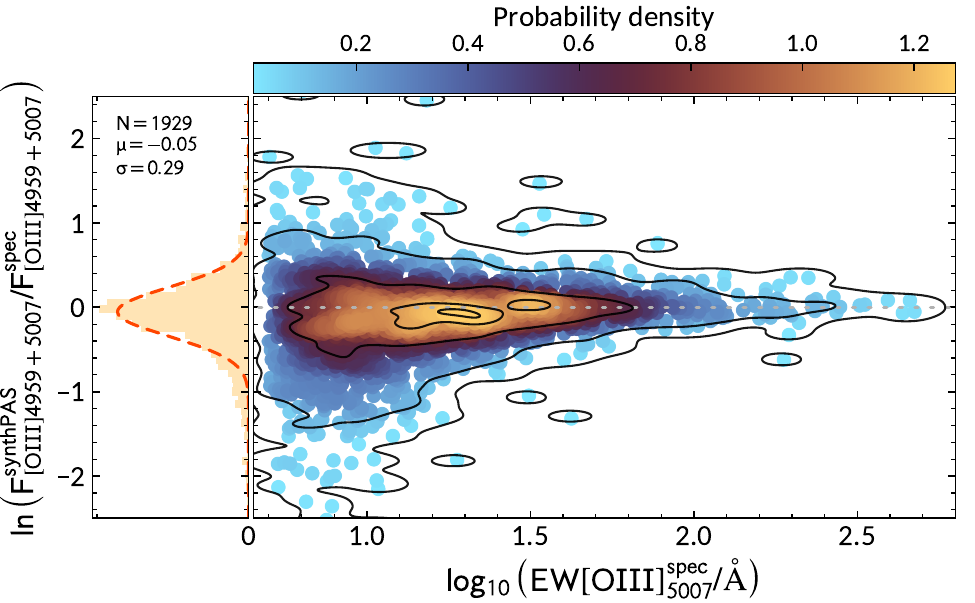}\label{subfig_desi_agn_ewo3}}~
  \subfigure[]{\includegraphics[width = 0.49\textwidth]{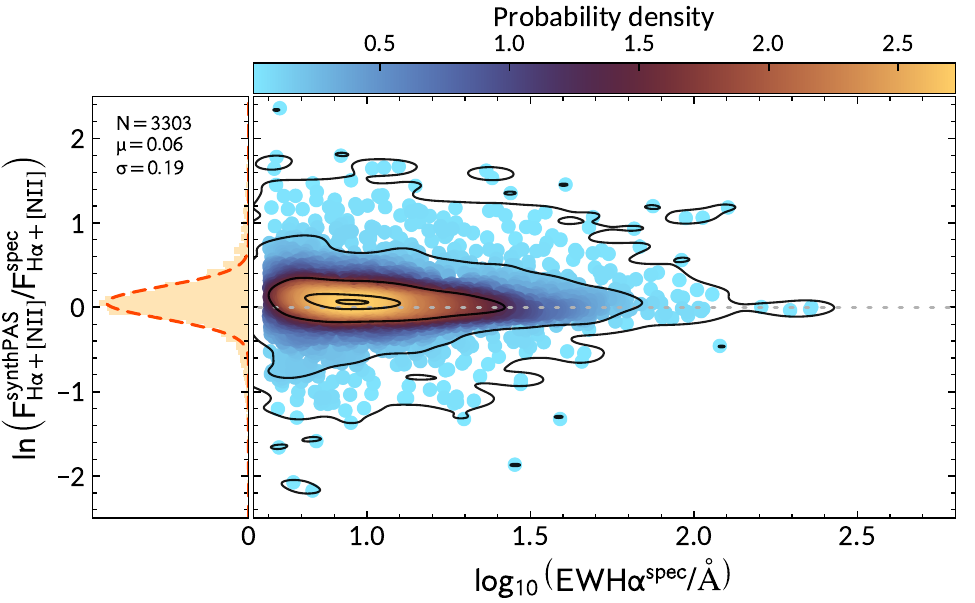}\label{subfig_desi_agn_ewha}}
  \caption{Same as Fig.~\ref{fig_desi}, but for the AGN population in the DESI EDR.}\label{fig_desi_agn}
  %The upper panels show the density plots of the emission-line fluxes for [\ion{O}{iii}]$\lambda 4959,5007$ (a) and H$\alpha$ + [\ion{N}{ii}]$\lambda 6548,6583$ (b) derived from the synthetic photometry of the AGN population in the DESI EDR \citep{desi24} spectra with the J-PAS filter system, using the SED fitting-based continuum subtraction method described in Section\,\ref{obs}. The J-PAS fluxes are compared with the spectroscopic line fluxes from the FastSpecFit value-added catalogue (\citealt{moustakas23}; Moustakas et al. in prep.). The lower panels show the line flux ratios between J-PAS line fluxes and the spectroscopic measurements for [\ion{O}{iii}]$\lambda 4959,5007$ (c) and H$\alpha$ + [\ion{N}{ii}]$\lambda 6548,6583$ (d), plotted as a function of the spectroscopic equivalent width of [\ion{O}{iii}]$\lambda 5007$ and H$\alpha$, respectively. In dashed grey style, we show the one-to-one lines in the upper panels and the constant unity value in the lower panels. The side vertical panels in (c) and (d) show the marginal distribution (grey histogram) and the corresponding Gaussian fit (dashed red line) of the J-PAS to spectroscopic line flux ratios. In all panels, the black contours represent the 1, 10, 50, 90, and 99\% percentiles of the two-dimensional probability distribution.
\end{figure*}

Following the methodology described in Section\,\ref{sec_desi}, the synthetic AGN photospectra were processed by performing continuum fitting with \textsc{CIGALE}, subtracting the best-fit continuum, and extracting the fluxes of the brightest emission lines listed in Table\,\ref{tab_lines}. The recovered emission-line fluxes were then directly compared with the corresponding spectroscopic measurements in the FastSpecFit catalogue (Fig.\,\ref{fig_desi_agn}). The derived line fluxes for AGN show excellent agreement with the spectroscopic values, confirming that robust continuum subtraction via SED fitting, without including an explicit AGN component, can still be applied successfully to this population.

This is primarily due to the moderate or low contribution of AGN to the optical continuum in these objects (note that bright, point-like quasars are excluded from the sample; see the target selection in Section\,\ref{obs}) and the flexibility of the extensive model grid (Table\,\ref{tab_cigpars}), which can accommodate potential AGN contamination in the continuum using stellar population templates. Moreover, the AGN templates available in \textsc{CIGALE} are primarily designed for modelling dust emission in the infrared and provide limited flexibility in the optical range, where the emission is typically represented by a single power-law. Therefore, including such models would not significantly enhance the continuum modelling in the optical regime.

This test confirms that our method for continuum subtraction and emission-line flux extraction can also be reliably applied to the AGN population, recovering line fluxes without introducing significant biases.

\end{appendix}
\end{document}